\newcommand{\diracslash}[1]{#1\llap{/\kern2pt}}

\newcommand{\be}{\begin{equation}}
\newcommand{\ee}{\end{equation}}
\newcommand{\bea}{\begin{eqnarray}}
\newcommand{\eea}{\end{eqnarray}}
\newcommand{\ba}[1]{\begin{array}{#1}}
\newcommand{\ea}{\end{array}}

\newcommand{\bt}{\begin{tabular}}
\newcommand{\et}{\end{tabular}}

\newcommand{\beas}{\begin{eqnarray*}}
\newcommand{\eeas}{\end{eqnarray*}}

\documentclass[preprint,prd,aps,amssymb,amsmath
,floats,nofootinbib,floatfix]{revtex4}

\DeclareSymbolFont{rsfs}{U}{rsfs}{m}{n}
\DeclareSymbolFontAlphabet{\mathrsfs}{rsfs}

\usepackage{graphicx}
\usepackage{multirow}
\usepackage{graphicx,epstopdf}

\usepackage{graphicx}
\usepackage{float}

\usepackage[utf8]{inputenc}
\usepackage[english]{babel}
\usepackage{hyperref}
\hypersetup{
    colorlinks=true,
    linkcolor=blue,
    filecolor=magenta,     
    citecolor=blue, 
    urlcolor=blue,
}
 
\usepackage{cleveref}

\begin{document}


\title{Heavy vector and axial-vector $D$ mesons in  hot  magnetized asymmetric nuclear matter}

\author{Rajesh Kumar}
\email{rajesh.sism@gmail.com}

\author{Rahul Chhabra}
\email{rahulchhabra@ymail.com}

\author{Arvind Kumar}
\email{iitd.arvind@gmail.com, kumara@nitj.ac.in}
\affiliation{Department of Physics, Dr. B R Ambedkar National Institute of Technology Jalandhar, 
 Jalandhar -- 144011,Punjab, India}
%

\def\be{\begin{equation}}
\def\ee{\end{equation}}
\def\bearr{\begin{eqnarray}}
\def\eearr{\end{eqnarray}}
\def\zbf#1{{\bf {#1}}}
\def\bfm#1{\mbox{\boldmath $#1$}}
\def\hf{\frac{1}{2}}
\def\kp{\zbf k+\frac{\zbf q}{2}}
\def\km{-\zbf k+\frac{\zbf q}{2}}
\def\hwo{\hat\omega_1}
\def\hwt{\hat\omega_2}

\begin{abstract}
We observed the impact of finite magnetic field on  the in-medium mass and  decay constant of isospin averaged vector $D^*(D^{*^+},D^{*^0})$ and axial-vector $D_1(D^+_1, D^0_1)$ mesons. The quark and gluon condensates of the nuclear medium at finite magnetic field, temperature, isospin asymmetry, and density  have been obtained by the meson exchange scalar fields within the chiral SU(3) model. The medium attributes modify the scalar and vector density of nuclear medium and this variation reflects in the in-medium mass and decay constant of spin 1 $D$ mesons. We  calculate these observables by comparing the Operator Product Expansion (OPE) and the phenomenological side in the   QCD Sum Rules. In the results, we observed a positive mass shift for charged vector and axial-vector $D$ mesons with respect to magnetic field. For neutral vector (axial-vector) $D$ mesons we observed negative (positive) mass shift as a function of magnetic field. In the application part, we calculate the in-medium partial decay width of the process  $D^*_s$(2715/2860) $\rightarrow$ $D^* K$ by using $^3P_0$ model. The in-medium effects are incorporated through the in-medium masses of $D^*$ and $K$ mesons. 

\end{abstract}

\maketitle

\maketitle

\section{Introduction}
\label{intro}

The in-medium study of $D$ meson is of great interest to understand the  $J/\psi$ suppression phenomenon which is considered as a signature of Quark Gluon Plasma (QGP) formation in Heavy Ion Collisions (HICs) \cite{Matsui1986}.  In addition to $J/\psi$ suppression, other indirect measurements such as jet quenching \cite{Bjorken1982}, dilepton enhancements \cite{Masera1995,Srivastava2009,Wilson1998} and strangeness enhancements \cite{Soff1999,Capella1995}  are used as a signature of QGP formation.  In $J/\psi$ suppression, due to color debye screening, the binding radius of charm quark pair becomes greater than the debye screening radius and hence the binding force cannot keep the charm and anticharm quark together \cite{Matsui1986,Chhabra2017}. The free charm quarks couples with the light quarks in the medium and  may form open charm mesons. The properties of $D$ (open charm meson) varies appreciably in the medium due to the presence of light quarks. Higher charmonia state decays to lower charmonium state ($J/\psi$ meson) directly, but if the mass of $D$ mesons decreases appreciably due to medium effects then it will prefer decaying into $D$ mesons and  eventually the $J/\psi$ get suppressed \cite{,Kumar2020,Chhabra2017,Matsui1986}. The in-medium properties of $D$ mesons varies considerably as the light quark condensates  varies appreciably in the medium \cite{Kumar2020}. On the other hand, in hidden charm/bottom mesons (charmonia/bottomonia) the properties do not vary appreciably as it depends upon the interaction of gluon condensates which do not change appreciably in the medium \cite{Kumar2020,Chhabra2018}.

Recently, in HICs the effect of the magnetic field was also found in addition to medium properties such as isospin asymmetry, temperature and density \cite{Kharzeev2008,Fukushima2008,Skokov2009}. The strength of generated magnetic field was calculated  $eB$ $\sim$ $2-15 m_{\pi}^2$  (1${{m}_{\pi}^2}$ = $ 2.818\times 10^{18}$ gauss) approximately \cite{Kumar2020}.  The presence of strong magnetic field urged physicists to find its impact on the phase transition, perturbative   and non-perturbative regime of physics \cite{Kharzeev2013,Fukushima2008,Vilenkin1980,
Burnier2011,Kumar2019,Kumar2020}. However, the time duration for which the magnetic field sustain is still unclear \cite{Reddy2018,Kumar2020}. According to the chiral magnetic effect, the magnetic field interacts with the nuclear matter and due to Lenz's law, it generates induced current, which affects the electric conductivity of the medium. The  change in electric conductivity increases the relaxation time that results in slow decay of magnetic field \cite{Kharzeev2013,Fukushima2008,Vilenkin1980,
Burnier2011}. The presence of the magnetic field also generates new interesting phenomenon like inverse magnetic catalysis, magnetic catalysis \cite{Kharzeev2013}. The effect of all these medium attributes on the properties of hadrons can be revealed in the upcoming Heavy Ion Collider (HIC), such as Compressed Baryonic Matter (CBM, GSI Germany), Japan Proton Accelerator Research Complex (J-PARC Japan),  Proton AntiProton Annihilation in Darmstadt (PANDA, GSI Germany) and Nuclotron-based Ion Collider Facility (NICA, Dubna Russia)  \cite{Rapp2010}.

Various  models have been constructed to study the non-perturbative regime of QCD. Some of these models are:   Nambu-Jona-Lasinio (NJL) model \cite{Nambu1961},   chiral SU(3) model \cite{Papazoglou1999,Mishra2004aa,Mishra2009a,Kumar2020,
Kumar2019}, Quark-Meson Coupling (QMC) model \cite{Guichon1988,Hong2001,Tsushima1999,Sibirtsev1999,Saito1994,Panda1997},     QCD sum rules (QCDSR)  \cite{Reinders1981,Hayashigaki2000,
Hilger2009,Reinders1985,Klingl1997,Klingl1999}, and  coupled channel approach \cite{Tolos2004,Tolos2006,Tolos2008,Hofmann2005}. In the above methodologies, the effect of quantum fluctuations are ignored by using  mean field potential. These quantum fluctuations can be  included by   the Polyakov loop extended NJL  (PNJL) model \cite{Fukushima2004,Kashiwa2008,Ghosh2015},  Polyakov Quark Meson (PQM) model \cite{Chatterjee2012,Schaefer2010}, and Functional Remormalization Group (FRG) \cite{Herbst2014,Drews2013}. Furthermore, the partial/full decay width of the heavy mesons have  been studied through several models such as  elementary meson-emission model \cite{Bonnaz1999},  $^3 S_1$ model \cite{Furui1987}, flux-tube model \cite{Kokoski1987} and $^3 P_0$ model \cite{Friman2002}.

The effect of magnetic field on the in-medium properties of vector $D^*$  and axial-vector $D_1$ mesons have not been studied in the literature. In the present work, we apply the combined approach of chiral SU(3) model and QCD sum rules to study the shift in mass and decay constant of these mesons in asymmetric magnetized hot nuclear matter.  
At first, using chiral model, we evaluate the magnetic field induced light quark condensates  $\left\langle \bar{q}q\right\rangle$,  and gluon condensates $\left\langle  \frac{\alpha_{s}}{\pi} {G^a}_{\mu\nu} {G^a}^{\mu\nu}
\right\rangle$, in dense and hot asymmetric nuclear matter. Afterwards, we input these condensates  in the Borel transformed QCD sum rules to calculate the in-medium  mass and decay constant for $D^*$ and $D_1$ mesons.  
In addition, as an application of in-medium mass shift of vector $D^*$ mesons, using $^3 P_0$ model we evaluate the in-medium  partial decay width of excited $D^*_s(2860)$ and $D^*_s(2715)$ mesons 
decaying to vector $D^*$ and pseudoscalar $K$ mesons.  The in-medium $K$ meson mass (at finite temperature, density and magnetic field)  for this purpose is calculated  using chiral SU(3) model.

 The effect of magnetic field on the in-medium properties of  hadrons has been studied by various methods \cite{Kumar2019,Kumar2019a,Kumar2020,Reddy2018,Gubler2016,
Mishra2019,Cho2014,Cho2015}. The properties of  charmonium \cite{Cho2015,Cho2014,Kumar2019,Kumar2019a}, bottomonium  \cite{Kumar2019a}, $\rho$ mesons \cite{Liu2015}, and $B$ mesons \cite{Machado2014,Dhale2018}, were studied in the presence of magnetic field. Using chiral SU(4) model, the author found an additional positive mass shift in the mass of $D^+$ meson under the influence of the magnetic field \cite{Reddy2018}. In Ref. \cite{Gubler2016}, using  QCD sum rules, Gubler $et. al.$  calculated the effect of the finite magnetic field on the mass of $D$ mesons as well as the mixing effect of vector and pseudoscalar $D$ mesons. The decay width of higher charmonia ($\psi(3770)$, $\psi(3686)$, ${\chi_c}_0(3414)$ and ${\chi_c}_1(3556)$) to $D \bar D$ pairs was also studied within the combined approach of $^3P_0$ model and chiral model \cite{Kumar2020}. In this article, we calculated the impact of magnetic field  on the mass and decay constant of pseudoscalar and scalar $D$ mesons using QCDSR and further used these properties to calculate in-medium partial decay width of excited charmonium states. In Ref. \cite{Kumar2019,Kumar2019a}, we also calculated the in-medium mass of charmonia and bottomonia in the presence of the magnetic field and found an attractive mass shift.  Extensive work is done on the in-medium effects without considering the magnetic field \cite{Kumar2010,Kumar2015,Kumar2014,Chhabra2017}.  For instance, unification of Operator Product Expansion (OPE)  and Borel transformation on the current-current correlation function was applied to  equate the mass and   various condensates \cite{Hayashigaki2000,Hilger2009}. The mass splitting between $D$ and $\bar{D}$ mesons was studied  using the linear density approximation in QCD sum rules \cite{Hilger2009}.
Using QCD sum rules and chiral SU(3) model, author evaluated the  shift in decay constants and mass of  vector and axial-vector  mesons \cite{Kumar2015}. In this article, they observed attractive interaction for vector $D^*$ and $B^*$ mesons whereas, repulsive interaction for axial-vector  $D_1$ and $B_1$ mesons  in hadronic medium. 

 The $^3P_0$ model had been  used to study the partial decay widths of different mesons \cite{Bruschini2019,Friman2002,close2007,sego2012,ferre12014,
ferre2014a,micu1969,yao1977,Zhang2007,barn2003,barn2005,close2005,Li2010}. 
It was predicted by the Babar and Belle collaboration that the excited states of $D_s$(1968), $i.e.$,   $D^*_s(2860)$ and  $D^*_s(2715)$ are of specific importance  having decay width  approximately  equal to 48 and 115 MeV, respectively. But, the spectroscopic state of $D^*_s$ meson is not yet confirmed. To find the exact quantum numbers, theoretical calculations have been done by observing different decay channels \cite{Zhang2007,Chhabra2017}. For instance, using  $^3P_0$ model, authors   investigated various decay channels of $D^*_s(2860)$ and  $D^*_s(2715)$  mesons and proposed the possible quantum numbers of $D^*_s(2860)$ and $D^*_s(2715)$ states such as  $3^-(1^3D_3)$,  $1^-(1^3D_1)$ and $1^-(2^3S_1)$ (least probable) \cite{liu2011,close2005,close2007,Godfrey2014,Zhang2007,
Chhabra2017}. By comparing the theoretical and empirical results, the study  of $D^*_s(2715$ or $2860)$ $\rightarrow$ $D^*$  + $K$ decay mode will help to reveal the possibility  to allocate $D^*_s$ meson quantum spectroscopic states \cite{Zhang2007,Chhabra2017}. The impact of magnetic field is studied on the $K$ and $\bar{K}$ mesons in cold asymmetric nuclear matter \cite{Mishra2019}. In this article, authors used chiral SU(3) model to calculate the in-medium mass of $K$ and $\bar{K}$ mesons and compared the results of zero and non zero contributions of anomalous magnetic moment. Moreover, the in-medium properties  of $K$ and  $\bar{K}$ mesons in zero magnetic field have  been studied extensively in the literature \cite{Mishra2004,Mishra2004a,Mishra2006,Mishra2008,Mishra2009}.

 The outline of the present paper is as follows: In subsection \ref{subsec:2.1}, we will briefly explain the methodology used to calculate the in-medium quark and gluon condensates and the mass of kaons in the nuclear medium. In subsection \ref{subsec:2.2}, we will explain the formalism to calculate the effective masses and decay constant of vector  and axial-vector $D$ mesons under the effect of magnetic field. The methodology to calculate the in-medium decay width of excited $D^*_s$ meson will be given in subsection \ref{subsec:2.3}. Section \ref{sec:3} will be devoted to discuss the quantitative results of the present work and finally, summary will be presented in section \ref{sec:4}.

\section{ Formalism}
\label{sec:2}

In the present work, we use the combination of two non-perturbative QCD techniques, (i) Chiral SU(3) model and (ii) QCD Sum Rules  \cite{Kumar2020,Hayashigaki2000}. From the first, we extract the magnetically induced light quark  and gluon condensates, which are used  in the QCD Sum Rules to evaluate the medium effects on the mass and decay constant of vector $D^*$ and axial-vector $D_1$ mesons. The medium modified  mass of kaons (calculated from chiral model) and $D^*$ meson are further used in the $^3P_0$ model to calculate the in-medium decay width of $D^*_s$ meson.  In the following subsections, we briefly describe the formulation to obtain the results. 

\subsection{In-medium quark and gluon condensates and mass of $K$ mesons}
\label{subsec:2.1}
 The chiral $SU(3)$ model includes the basic QCD properties such as  trace anomaly (broken scale invariance) and non-linear realization of chiral symmetry \cite{Weinberg1968,Coleman1969,Zschiesche1997,Bardeen1969,
Kumar2020,Papazoglou1999,Kumar2019}. In this model, the isospin asymmetry of the medium is introduced  by  including the scalar isovector field $\delta$ and vector-isovector field $\rho$ \cite{Kumar2020}.  Moreover, the broken scale invariance property  is preserved by the inclusion of scalar dilaton field $\chi$ \cite{Papazoglou1999,Kumar2020}. The effect of thermal and quantum fluctuations are neglected in the present model by using mean field approximation \cite{Kumar2020,Reddy2018}. The effect of the external magnetic field is incorporated by adding the magnetic  induced Lagrangian density  to the chiral model's Lagrangian density \cite{Kumar2019,Reddy2018}.   The coupled equations of motion  of the  chiral model fields are solved by minimizing the effective thermodynamic potential of chiral SU(3) model \cite{Kumar2019,Kumar2019a} and are given as

\begin{eqnarray}
 k_{0}\chi^{2}\sigma-4k_{1}\left( \sigma^{2}+\zeta^{2}
+\delta^{2}\right)\sigma-2k_{2}\left( \sigma^{3}+3\sigma\delta^{2}\right)
-2k_{3}\chi\sigma\zeta \nonumber\\
-\frac{d}{3} \chi^{4} \bigg (\frac{2\sigma}{\sigma^{2}-\delta^{2}}\bigg )
+\left( \frac{\chi}{\chi_{0}}\right) ^{2}m_{\pi}^{2}f_{\pi}
=\sum g_{\sigma i}\rho_{i}^{s} ,
\label{sigma}
\end{eqnarray}
\begin{eqnarray}
 k_{0}\chi^{2}\zeta-4k_{1}\left( \sigma^{2}+\zeta^{2}+\delta^{2}\right)
\zeta-4k_{2}\zeta^{3}-k_{3}\chi\left( \sigma^{2}-\delta^{2}\right)\nonumber\\
-\frac{d}{3}\frac{\chi^{4}}{\zeta}+\left(\frac{\chi}{\chi_{0}} \right)
^{2}\left[ \sqrt{2}m_{K}^{2}f_{K}-\frac{1}{\sqrt{2}} m_{\pi}^{2}f_{\pi}\right]
 =\sum g_{\zeta i}\rho_{i}^{s} ,
\label{zeta}
\end{eqnarray}
\begin{eqnarray}
k_{0}\chi^{2}\delta-4k_{1}\left( \sigma^{2}+\zeta^{2}+\delta^{2}\right)
\delta-2k_{2}\left( \delta^{3}+3\sigma^{2}\delta\right) +2k_{3}\chi\delta
\zeta \nonumber\\
 +   \frac{2}{3} d \chi^4 \left( \frac{\delta}{\sigma^{2}-\delta^{2}}\right)
=\sum g_{\delta i}\tau_3\rho_{i}^{s}  ,
\label{delta}
\end{eqnarray}

\begin{eqnarray}
\left (\frac{\chi}{\chi_{0}}\right) ^{2}m_{\omega}^{2}\omega+g_{4}\left(4{\omega}^{3}+12{\rho}^2{\omega}\right) =\sum g_{\omega i}\rho_{i}^{v}  ,
\label{omega}
\end{eqnarray}

\begin{eqnarray}
\left (\frac{\chi}{\chi_{0}}\right) ^{2}m_{\rho}^{2}\rho+g_{4}\left(4{\rho}^{3}+12{\omega}^2{\rho}\right)=\sum g_{\rho i}\tau_3\rho_{i}^{v}  ,
\label{rho}
\end{eqnarray}

and

\begin{eqnarray}
k_{0}\chi \left( \sigma^{2}+\zeta^{2}+\delta^{2}\right)-k_{3}
\left( \sigma^{2}-\delta^{2}\right)\zeta + \chi^{3}\left[1
+{\rm {ln}}\left( \frac{\chi^{4}}{\chi_{0}^{4}}\right)  \right]
+(4k_{4}-d)\chi^{3}
\nonumber\\
-\frac{4}{3} d \chi^{3} {\rm {ln}} \Bigg ( \bigg (\frac{\left( \sigma^{2}
-\delta^{2}\right) \zeta}{\sigma_{0}^{2}\zeta_{0}} \bigg )
\bigg (\frac{\chi}{\chi_0}\bigg)^3 \Bigg )+
\frac{2\chi}{\chi_{0}^{2}}\left[ m_{\pi}^{2}
f_{\pi}\sigma +\left(\sqrt{2}m_{K}^{2}f_{K}-\frac{1}{\sqrt{2}}
m_{\pi}^{2}f_{\pi} \right) \zeta\right] \nonumber\\
-\frac{\chi}{{{\chi_0}^2}}(m_{\omega}^{2} \omega^2+m_{\rho}^{2}\rho^2)  = 0 ,
\label{chi}
\end{eqnarray}

respectively.

In above,  the constants  $m_\pi$, $m_K$, $f_\pi$ and $f_K$  are the masses and decay constants of pions and kaons, respectively and the parameters $k_i(i=1$ to $4)$ are fitted to reproduce the vacuum values of scalar meson fields \cite{Kumar2010}. In addition,  the  isospin asymmetry is incorporated in the model through definition, $\eta = -\frac{\Sigma_i \tau_{3i} \rho^{v}_{i}}{2\rho_{N}}$. Where $\rho^{v}_{i}$ and $\rho^{s}_{i}$ denote the  vector and scalar densities of $i^{th}$ nucleons ($i=p,n$)  in the presence of magnetic field (along $Z$-direction) \cite{Kumar2019,Broderick2000,Broderick2002}. 

In this model, through the explicit symmetry breaking terms, the scalar up and down quark condensates  are expressed as \cite{Kumar2020}


\begin{align}
\left\langle \bar{u}u\right\rangle_{\rho_N}
= \frac{1}{m_{u}}\left( \frac {\chi}{\chi_{0}}\right)^{2}
\left[ \frac{1}{2} m_{\pi}^{2}
f_{\pi} \left( \sigma + \delta \right) \right],
\label{qu}
\end{align}

and

\begin{align}
\left\langle \bar{d}d\right\rangle_{\rho_N}
= \frac{1}{m_{d}}\left( \frac {\chi}{\chi_{0}}\right)^{2}
\left[ \frac{1}{2} m_{\pi}^{2}
f_{\pi} \left( \sigma - \delta \right) \right],
\label{qd}
\end{align}

respectively. Here, $m_u$ and $m_d$ are the masses of up and  down quarks respectively.
Furthermore, by using the broken scale invariance property of QCD \cite{Papazoglou1999,Kumar2020}, the scalar gluon condensate ${G_0}_{\rho_N}$=$\left\langle \frac{\alpha_{s}}{\pi} 
G^a_{\mu \nu} {G^a}^{\mu \nu} \right\rangle_{\rho_N} $ is derived by the comparison of energy-momentum tensor of chiral model  with the energy-momentum tensor of QCD. It is expressed  in terms of  scalar fields as \cite{Kumar2019}

\begin{eqnarray}
\left\langle \frac{\alpha_{s}}{\pi} 
G^a_{\mu \nu} {G^a}^{\mu \nu} \right\rangle_{\rho_N} =  \frac{8}{9} \Bigg [(1 - d) \chi^{4}
+\left( \frac {\chi}{\chi_{0}}\right)^{2} 
\left( m_{\pi}^{2} f_{\pi} \sigma
+ \big( \sqrt {2} m_{K}^{2}f_{K} - \frac {1}{\sqrt {2}} 
m_{\pi}^{2} f_{\pi} \big) \zeta \right) \Bigg ].
\label{chiglum}
\end{eqnarray}

Here, we have taken  $d$=0.1818  from QCD beta function, $\beta_{QCD}$  at the one loop level \cite{Kumar2010}. 

As said before, to calculate the in-medium decay width of $D^*_s$ meson, we need in-medium mass of $K$ mesons. To obtain the mass modification of $K (\bar K)$ meson, we start from kaon-antikaon interaction Lagrangian density \cite{Mishra2019}
\begin{eqnarray}
\mathcal{L}_{K}^{int}&=&-\frac{i}{4f_k^2}[(2\bar{p}\gamma^\mu p
+\bar{n}\gamma^\mu n)
(K^-(\partial_ \mu K^+)
-(\partial _ \mu K^-)K^+)
\nonumber \\ &+& (\bar{p}\gamma^\mu p+2\bar{n}\gamma^\mu n)
(\bar{K^0}(\partial_\mu K^0)
-(\partial_\mu\bar{K^0})K^0)]
\nonumber \\ &+& \frac{m_K^2}{2f_k^2}
\Big [ (\sigma+\sqrt{2}\zeta+\delta)(K^+K^-)
+  (\sigma+\sqrt{2}\zeta-\delta)(K^0\bar{K^0})\Big]\nonumber\\&-
&\frac{1}{f_k}[(\sigma+\sqrt{2}\zeta+\delta)(\partial_\mu K^+)
(\partial^ \mu {K^-})+(\sigma+\sqrt{2}\zeta-\delta)
(\partial^ \mu K^0)(\partial^ \mu \bar{K^0})]\nonumber\\&+
&\frac{d_1}{2f_k^2}[(\bar{p}p+\bar{n}n)((\partial_\mu K^+)
(\partial^ \mu K^-)+(\partial_\mu K^0)
(\partial^ \mu \bar{K^0}))]
\nonumber\\&+&\frac{d_2}{2f_k^2}[\bar{p}p
(\partial_\mu K^+)(\partial^ \mu K^-)
+\bar{n}n(\partial_\mu K^0)(\partial^ \mu \bar{K^0})].
\label{lk_int}
\end{eqnarray}
The first term in the above expression is the vectorial Weinberg Tomozawa interaction term. This term have leading order contribution and is  repulsive for $K$ mesons but attractive for $\bar K$
mesons \cite{Mishra2019}. Further, the scalar meson exchange term represents the next to leading order contributions. The parameters $d_1$ and $d_2$ in the above expression
are taken as $ 2.56/m_K $ and $ 0.73/m_K $ 
respectively \cite{Mishra2019}, 
by fitting the experimental values 
of the kaon-nucleon ($KN$) scattering length \cite{Barnes1994}. 

The total Lagrangian density for kaon-antikaon is given by

\begin{equation}
{\cal L}_K =(\partial _\mu {\bar K}) (\partial^\mu K)-
m_{K(\bar K)}^2 \bar K K +{\cal L}_K^{int}.
\end{equation}
Now, by performing Fourier transformation on the above expression, we get following dispersion relation for kaons

\begin{equation}
-\omega^2 + |{\vec k}|^2 +m_{K}^2
-\Pi_{K}(\omega, | \vec{k} | )=0.
\label{dispkkbar}
\end{equation}
In above,  $\Pi_{K}$ represents the kaon  in-medium self energy and is explicitly given  as

\begin{eqnarray}
&&\Pi_{K}(\omega,|\vec{k}|)=-\frac{1}{4f_K^2}
[3(\rho_p+\rho_n)\pm(\rho_p-\rho_n)] \omega 
+\frac{m_K^2}{2f_K}(\sigma^\prime+\sqrt{2}\zeta^\prime\pm\delta^\prime)
\nonumber \\ &+&
\Bigg[-\frac{1}{f_K}(\sigma^\prime+\sqrt{2}\zeta^\prime\pm\delta^\prime)
+\frac{d_1}{2f_K^2}(\rho_p^s+\rho_n^s) 
+\frac{d_2}{4f_K^2}[(\rho_p^s+\rho_n^s)
\pm(\rho_p^s-\rho_n^s)]\Bigg](\omega^2- |{\vec k}|^2)\nonumber,\\
\label{selfk}
\end{eqnarray}
where the $\pm$ signs represents the  $K^+$ and $K^0$ mesons, respectively. Also, in the above expression  $\sigma '$, $\zeta '$ and $\delta '$   denote the  deviation of the field's expectation values 
from their vacuum expectation values. In magnetized nuclear matter, the masses for $K$ meson is calculated under the condition
$m_{K}^*=\omega(|\vec k|$=0)). The charged $K^+$ meson in magnetic medium gets additional positive mass shift due to the Landau quantization \cite{Kumar2020}
\begin{eqnarray}
m_{K^{+}}^{**}=\sqrt{m_{K^{+}}^*{^2}+eB},
\label{meffkpkm}
\end{eqnarray}
where $m_{K^+}^*$ is the in-medium mass of  $K^+$ meson, $e$ is the electrostatic unit of charge and $B$ is the magnetic field along $Z$-direction. On the other hand, 
for uncharged $K^0$ meson there  will not be Landau quantization.

\subsection{In-medium mass and decay constant of vector and axial-vector $D$ mesons}
\label{subsec:2.2}

QCD Sum Rules are low energy QCD methodology to relate the hadron properties from medium to vacuum  \cite{Reinders1985,Hayashigaki2000}. These rules are based on the Operator Product Expansion and Borel transformation to overcome the divergent behaviour of the perturbative expansion \cite{Reinders1985,Klingl1999}. In these rules, we have taken into account the next to leading order contributions in the Borel transformed coefficients. We start with a current current correlator function, which is a Fourier transformation of time ordered product of the meson current, $J(x)$ at some value of nuclear density, $\rho_N$ and temperature, $T$ \cite{Kumar2020,Wang2015}


\begin{align} \label{pb2}
\Pi(p) = i\int d^{4}x\ e^{ip. x} \langle \mathcal{T}\left\{J(x)J^{\dag}(0)\right\} \rangle_{\rho_N,T},
\end{align}
where $p$ is the four momentum. In the present work, we have used centroid approximation, in which we have used same rules (degeneracy) for particles and antiparticles. The mass splitting between particle and antiparticle can be studied by considering even and odd sum rules \cite{Hilger2009}.  The isospin average meson currents of the degenerate  vector and axial-vector $D$ mesons are given by the relations
\begin{eqnarray}
 J_\mu(x) &=&J_\mu^\dag(x) =\frac{\bar{c}(x)\gamma_\mu q(x)+\bar{q}(x)\gamma_\mu c(x)}{2},
    \end{eqnarray}
 and
\begin{eqnarray}
  J_{5\mu}(x) &=&J_{5\mu}^\dag(x) =\frac{\bar{c}(x)\gamma_\mu \gamma_5q(x)+\bar{q}(x)\gamma_\mu\gamma_5 c(x)}{2},
\end{eqnarray}
 respectively. Here, the quark operator, $q(x)$ represents the respective light quark content of $D^*$($D_1$) mesons whereas $c(x)$ represents the charm quark operator. The  quark composition of $D^*{^+}$, $D^*{^0}$, $D^+_1$ and $D^0_1$ mesons are   $c\bar d$, $c\bar u$, $c\bar d$ and $c\bar u$ respectively.  The mass splitting of the  vector  $D^*$($D^*{^+}$, $D^*{^0}$) and axial-vector $D_1$($D^+_1$,$D^0_1$) isospin doublet will be studied under the effect of isospin asymmetric matter. However, we will see the splitting will not be much visible due to the interference of magnetic field. We can divide current-current correlator in three parts (i) vacuum (ii) static nucleon and (iii) pion bath thermal, $i.e.,$

 \begin{align}
\Pi (p) =\Pi_{0} (p)+ \frac{\rho_{N}}{2m_N}T_{N} (p) + \Pi_{P.B.}(p,T).
\label{pb}
 \end{align}
 
In above, $m_N$ denotes the vacuum mass of nucleons  and  $T_N (p)$ denotes the forward scattering amplitude.  In the QCDSR, the pion bath term is used to incorporate the temperature effects of the medium, but in the present approach, we will introduce the finite temperature effects by the quark and gluon condensates which are calculated through the scalar fields within the chiral model. These meson fields $\sigma$, $\zeta$, $\delta$ and $\chi$ are solved as a coupled equations under the effect of temperature, magnetic field, density and isospin asymmetry \cite{Kumar2019}.  Therefore, we neglect the contribution of the thermal term in the correlator function and write
 
 \begin{eqnarray}
\Pi(p) =\Pi_{0}(p)+ \frac{\rho_N}{2m_N}T_{N}(p)\, .
 \end{eqnarray}

In the limit of $p\rightarrow$0, the scattering amplitude, $T_N (p)$ is represented in terms of spectral density which is parametrized in three unknown parameters $a$, $b$, and $c$. From the expression of the phenomenological spectral density and forward scattering amplitude, one can find the scattering length which is given as \cite{Wang2015}

\begin{eqnarray}
a_{D^*/D_1}=\frac{a}{f_{D^*/D_1}^2m_{D^*/D_1}^2(-8\pi(m_N+m_{D^*/D_1}))}.
\end{eqnarray}

To get the values of unknown parameters $a$ and $b$, we equate the Borel transformed scattering matrix of OPE side with the Borel transformed scattering matrix on the phenomenological side \cite{Hayashigaki2000}. By doing this, we get a mathematical relation  in the Borel transformed coefficients and unknown parameters $a$  and $b$, which is
\begin{eqnarray}
 a\, C_a+b\, C_b &=&C_f \, .
 \label{wdiff}
\end{eqnarray}
The explicit form of Borel transformed coefficients $C_a, C_b$ and $C_f$ for vector meson current,  $J_\mu(x)$  upto next to leading order contributions are given as \cite{Wang2015},
\begin{eqnarray}
C_a &=&\frac{1}{M^2}\exp\left(-\frac{m_{D^*}^2}{M^2}\right)-\frac{s_0}{m_{D^*}^4}\exp\left(-\frac{s_0}{M^2}\right) \, ,\nonumber\\
C_b&=&\exp\left(-\frac{m_{D^*}^2}{M^2}\right)-\frac{s_0}{m_{D^*}^2}\exp\left(-\frac{s_0}{M^2}\right) \, ,
\end{eqnarray}
\begin{eqnarray}
C_f&=& \frac{2m_N(m_H+m_N)}{(m_H+m_N)^2-m_{D^*}^2}\left(f_{D^*}m_{D^*}g_{D^*NH}\right)^2\left\{ \left[\frac{1}{M^2}-\frac{1}{m_{D^*}^2-(m_H+m_N)^2}\right] \exp\left(-\frac{m_{D^*}^2}{M^2}\right)\right.\nonumber\\
&&\left.+\frac{1}{(m_H+m_N)^2-m_{D^*}^2}\exp\left(-\frac{(m_H+m_N)^2}{M^2}\right)\right\}-\frac{m_c\langle\bar{q}q\rangle_N}{2}\left\{1+\frac{\alpha_s}{\pi} \left[ \frac{8}{3}-\frac{4m_c^2}{3M^2} \right.\right.\nonumber\\
&&\left.\left.+\frac{2}{3}\left( 2+\frac{m_c^2}{M^2}\right)\log\frac{m_c^2}{\mu^2}-\frac{2m_c^2}{3M^2}\Gamma\left(0,\frac{m_c^2}{M^2}\right)\exp\left( \frac{m_c^2}{M^2}\right) \right]\right\}\exp\left(- \frac{m_c^2}{M^2}\right) \nonumber\\
&&+\frac{1}{2}\left\{-\frac{4\langle q^\dag i D^*q\rangle_N}{3} +\frac{2m_c^2\langle q^\dag i D^*q\rangle_N}{M^2}+\frac{2m_c\langle\bar{q}g_s\sigma Gq\rangle_N}{3M^2}+\frac{16m_c\langle \bar{q} i D^* i D^*q\rangle_N}{3M^2}\right.\nonumber\\
&&\left.-\frac{2m_c^3\langle \bar{q} i D^* i D^*q\rangle_N}{M^4}-\frac{1}{12}\langle\frac{\alpha_sGG}{\pi}\rangle_N\right\}\exp\left(- \frac{m_c^2}{M^2}\right)\, ,
\label{coef}
\end{eqnarray}
 and for the axial-vector current $J_{5\mu}(x)$, the following transformations are used in the vector current,  where $i=a,b,f$
\begin{eqnarray}
 C_i &\to& C_i\left( {\rm with}\,\,\, m_N \to -m_N\, , \,\,\,m_c \to -m_c\, , \,\,\,D^* \to D_1\right)\, .
 \end{eqnarray}

In above equations,  the $\langle \bar q q \rangle_N$, $\langle q^\dag i D q\rangle_N$, $\langle \bar{q} i D i Dq\rangle_{_N}$ and  $ \langle \frac{\alpha_s GG}{\pi} \rangle_N$ are the nucleon expectation values of quark and gluon condensates, respectively. The fraction $\frac{1}{M^2}$ is the Borel operator.  The  in-medium quark and gluon condensates can be expressed in the nucleon expectation via relation \cite{Chhabra2017,Kumar2020}
\begin{eqnarray}
 \langle{\cal{O}}\rangle_{N} &= &\frac{2m_N}{\rho_N}\left( \langle{\cal{O}}\rangle_{\rho_N}-\langle {\cal{O}}\rangle_{vac}\right),
\end{eqnarray}
where $\cal{O}$ represents any of the four condensate,  $\langle{\cal{O}}\rangle_{\rho_N}$ denotes the medium dependent expectation value and  the   $\langle{\cal{O}}\rangle_{vac}$  denotes the vacuum expectation value. The condensates $\langle \bar u u \rangle_{\rho_N}$,  $\langle \bar d d \rangle_{\rho_N}$ and  $ \langle \frac{\alpha_s GG}{\pi} \rangle_{\rho_N}$ are calculated from the chiral SU(3) model and are given in Eqs. (\ref{qu}), (\ref{qd}) and (\ref{chiglum}), respectively.  The value of condensate $\langle \bar{q} i D i Dq\rangle_{\rho_N}$ and $\langle q^\dag i D q\rangle_N$ is used  from the linear density approximations results \cite{Thomas2007,Kumar2020}.

 Now, in Eq.(\ref{wdiff}), we have two unknown parameters but one equation. To solve this, we need one more equation. To achieve this, we differentiate the first equation by $z=\frac{1}{M^2}$. Therefore, the modified equation becomes 

\begin{eqnarray}
 a\, \frac{d}{dz} C_a+b\, \frac{d}{dz}C_b &=&\frac{d}{dz}C_f \,.
 \label{diff}
\end{eqnarray}
From Eqs.(\ref{wdiff}) and (\ref{diff}), we can write
 \begin{eqnarray}
 a&=&\frac{C_f\left(-\frac{d}{dz}\right)C_b-C_b\left(-\frac{d}{dz}\right)C_f}{C_a\left(-\frac{d}{dz}\right)C_b-C_b\left(-\frac{d}{dz}\right)C_a}\, , \nonumber\\
  b&=&\frac{C_f\left(-\frac{d}{dz}\right)C_a-C_a\left(-\frac{d}{dz}\right)C_f}{C_b\left(-\frac{d}{dz}\right)C_a-C_a\left(-\frac{d}{dz}\right)C_b}\, .
 \end{eqnarray}

The in-medium mass shift and shift in decay constant can be expressed in terms of unknown parameters $a$ and $b$ via   relations \cite{Wang2015}

\begin{eqnarray}
{\Delta m}^*_{D^*/D_1} &=&2\pi\frac{(m_{N}+m_{D^*/D_1})a\rho_N}{m_Nm_{D^*/D_1}f_{D^*/D_1}^2m_{D^*/D_1}^2(-8\pi(m_N+m_{D^*/D_1}))}\, ,
\label{msd}
\end{eqnarray}

and

\begin{eqnarray}
\Delta f^*_{D^*/D_1}&=&\frac{1}{2f_{D^*/D_1}m_{D^*/D_1}^2}\left(\frac{b\rho_N}{2m_N}-2f_{D^*/D_1}^2m_{D^*/D_1}\Delta m^*_{D^*/D_1} \right) \, , \nonumber\\.
\label{fd}
\end{eqnarray}

respectively.  The in-medium mass of $D$ mesons can be written as 
\begin{equation}
m^{*}_{D^*/D_1}=m_{D^*/D_1}+\Delta m_{D^*/D_1}.
\end{equation} 

Here, $m_{D^*/D_1}$ denotes vacuum mass of vector and axial-vector $D$ mesons. The in-medium mass of charged $D^*{^+}$ and $D^+_1$ mesons experiences an additional positive shift in mass through the interaction of charge particle with magnetic field, and this effect is called Landau quantization. The modified expression for mass is given as

\begin{equation}
m^{**}_{D^{*^+},D^+_1}=\sqrt {{m^*_{D^{*^+},D^+_1}}^2 +|eB|}.
\label{mdpm_landau}
\end{equation}

As was the case for neutral $K$ meson, neutral vector and axial-vector $D$ meson  does not experience any additional mass shift due to Landau quantization.

%
%
%
%

%

\subsection{In-medium decay width of $D^*_s$ $\rightarrow$ $D^*$ $K$  meson using $^3P_0$ model} 
\label{subsec:2.3}
We use $^3 P_0$ model to study the effect of  mass shift of $D^*$ and $K$ meson on the   partial decay widths of $D^*_s(2715)$ and $D^*_s(2860)$ states decaying to $D^*$ and $K$ mesons \cite{Zhang2007,Chhabra2017}. As discussed earlier,  the parent meson $D^*_s$  is proposed to be in  the  $3^-(1^3D_3)$,  $1^-(1^3D_1)$ and $1^-(2^3S_1)$  spectroscopic states \cite{liu2011,close2005,close2007,Godfrey2014,Zhang2007,
Chhabra2017}.  In this model, we assume the creation of quark and anti-quark pair having quantum numbers $0^{++}$. The helicity amplitude is given by \cite{Zhang2007} 
\begin{align}\label{M}
 \mathcal{M}^{M_{J_{D^*_s} } M_{J_{D^*} } M_{J_K }} = \gamma  \sqrt {8E_{D^*_s} E_{D^*} E_K } \sum_{\substack{M_{L_{D^*_s} } ,M_{S_{D^*_s} } ,M_{L_{D^*} }, \\M_{S_{D^*} } ,M_{L_K} ,M_{S_K } ,m} }\langle {1m;1 - m}|{00} \rangle \nonumber \\
 \times \langle {L_{D^*_s} M_{L_{D^*_s} } S_{D^*_s} M_{S_{D^*_s} } }| {J_{D^*_s} M_{J_{D^*_s} } }\rangle \langle L_{D^*} M_{L_{D^*} } S_{D^*} M_{S_{D^*} }|J_{D^*} M_{J_{D^*} } \rangle\langle L_K M_{L_K } S_K M_{S_K }|J_K M_{J_K }\rangle \nonumber \\
  \times\langle\varphi _{D^*}^{13} \varphi _K^{24}|\varphi _{D^*_s}^{12}\varphi _0^{34} \rangle
\langle \chi _{S_{D^*} M_{S_{D^*} }}^{13} \chi _{S_K M_{S_K } }^{24}|\chi _{S_{D^*_s} M_{S_{D^*_s} } }^{12} \chi _{1 - m}^{34}\rangle I_{M_{L_{D^*} } ,M_{L_K } }^{M_{L_{K}} ,m} (\textbf{k}).
\end{align}
In above,  $E_{D^*_s}$= $m_{D^*_s}$, $E_{D^*}$ = $\sqrt{m_{D^*}^{*2} + k_{D^*}^2}$ and $E_{K}$ = $\sqrt{m_{K}^{*2} + k_K^2}$ represent the effective energies of respective mesons. Here $m_{D^*}^*$ and $m^*_K$ are the in-medium masses of $D^*$ and $K$ mesons respectively. The partial wave amplitude is obtained by transformation of helicity amplitude using Jacob-Wick formula \cite{Zhang2007,Chhabra2017}

\begin{eqnarray}
\mathcal{M}^{JL} (D^*_s \to D^* K) &=& \frac{{\sqrt {2{L} + 1} }}{{2{ J_{D^*_s}} + 1}}\sum_{M_{J_{D^*}},M_{ J_K}} \langle{{ L}0{J} M_{J_{D^*_s}}} |{J_{D^*_s} M_{J_{D^*_s} } }\rangle \nonumber \\
&&\times \left\langle {{ J_{D^*}} M_{J_{D^*}} { J_K} M_{J_K}}\right|\left. {{J} M_{J_{D^*_s} } } \right\rangle M^{M_{J_{D^*_s} } M_{J_{D^*}} M_{J_K } } (\textbf{$k_{D^*}$}).
\end{eqnarray}

%
%
%
%
%
%

In above, $M_{J_{D^*_s}}=M_{J_{D^*}}+M_{J_K}$, $|{{J_{D^*}}}-{{J_K}}| \leq $ ${{J}}$  $\leq$ $|{{J_{D^*}}} +{ J_K}|$  and $|{J} - {L}| \leq { J_{D^*_s}} $  $\leq |{ J} + { L}|$.

The decay width of $D^*_s$ meson can be calculated by the following formula
\begin{align}
\Gamma  =  \frac{|{\textbf{$k_{D^*}$}}|}{8\pi m_{D^*_s}^2}\sum_{JL} |{\mathcal{M}^{JL}}|^2.
\label{G}
\end{align}
Since the decaying meson is assumed to be at rest, the magnitude of the  momentum of $D^*$ and $K$ meson is same,$i.e.$, $|k_{D^*}|=|k_K|$ and is given as
\begin{align}
 |\textbf{$k_{D^*}$}|= \frac{{\sqrt {[m_{D^*_s}^2  - (m^*_{D^*}  - m^*_K )^2 ][m_{D^*_s}^2  - (m^*_{D^*} + m^*_K )^2 ]} }}{{2m_{D^*_s} }}.
\label{K}
\end{align}

 Thus, the in-medium partial decay widths of different spectroscopic states of $D^*_s(2715)$ and $D^*_s(2860)$ states decaying to ($D^*$, $K$) can be calculated.  
 

\subsubsection{$\mathbf{1^{-}(1^{3}D_{1})}$ $\mathbf{c\bar{s}}$\textbf{ state}}

The harmonic-oscillator wave function of $1^{-}(1^{3}D_{1})$
$c\bar{s}$ state will be
\begin{eqnarray} \phi^{n=1;L=2}({{\mathbf{k}_1,\mathbf{k}_2}}) &=& \frac{R^{7/2}}{\sqrt{15}\pi^{1/4}}
\mathcal{Y}_{2}^{m}\big(\frac{\mathbf{k}_{1}-\mathbf{k}_{2}}{2}\big)
\exp\Big{[}-{1\over 8}({{\mathbf{k}_1-\mathbf{k}_2})}^2R^2\Big{]},
\end{eqnarray}
where $R$ denotes the radius of the meson. We have the decay matrix amplitude \cite{Zhang2007}
\begin{eqnarray}
\mathcal{M}\big(c\bar{s}(1^{3}D_{1})\to
1^{-}+0^{-}\big)=\alpha\frac{\gamma\sqrt{8E_{D^*_s}E_{D^*}E_{K}}}{\sqrt{18}}\big[\frac{1}{\sqrt{30}}I^{0,0}
+\frac{1}{\sqrt{40}}I^{1,-1}+\frac{1}{\sqrt{40}}I^{-1,1}\big],
\end{eqnarray}

where
\begin{eqnarray}
I^{0,0}&=\frac{|\mathbf{k}_{D^*}|\pi^{1/4}R_{D^*_s}^{7/2}R_{D^*}^{3/2}R_{K}^{3/2}(\Upsilon-1)}{(R_{D^*_s}^{2}+R_{D^*}^{2}+R_{K}^{2})^{5/2}}\bigg[
(R_{D^*_s}^{2}+R_{D^*}^{2}+R_{K}^{2})(\Upsilon^{2}-1)\mathbf{k}_{D^*}^{2}+8\bigg]\nonumber\\
&\exp\bigg[-\frac{\mathbf{k}_{D^*}^{2}R_{D^*_s}^{2}(R_{D^*}^{2}+R_{K}^{2})}{8(R_{D^*_s}^{2}+R_{D^*}^{2}+R_{K}^{2})}\bigg],\label{I00}\\
I^{1,-1}&=I^{-1,1}=-\frac{4\sqrt{3}|\mathbf{k}_{D^*}|\pi^{1/4}R_{D^*_s}^{7/2}R_{D^*}^{3/2}R_{K}^{3/2}(\Upsilon-1)}{(R_{D^*_s}^{2}
+R_{D^*}^{2}+R_{K}^{2})^{5/2}}\exp\bigg[-\frac{\mathbf{k}_{D^*}^{2}R_{D^*_s}^{2}(R_{D^*}^{2}+R_{K}^{2})}
{8(R_{D^*_s}^{2}+R_{D^*}^{2}+R_{K}^{2})}\bigg]\label{I11}.
\end{eqnarray}
The symbols $D^*_s$, $D^*$ and $K$ denote $1^{-}(1^{3}D_{1})$ 
$c\bar{s}$ state, vector $D$ meson and pseudoscalar kaon, respectively.

\subsubsection{$\mathbf{3^{-}(1^{3}D_{3})}$ $\mathbf{c\bar{s}}$\textbf{ state}}

The general decay matrix amplitude for $c\bar{s}(1^{3}D_{3})\to
1^{-}+0^{-}$ can be written as
\begin{eqnarray}
\mathcal{M}\big(c\bar{s}(1^{3}D_{3})\to
1^{-}+0^{-}\big)=\alpha\frac{\gamma\sqrt{8E_{D^*_s}E_{D^*}E_{K}}}{\sqrt{42}}\big[\frac{2}{\sqrt{30}}I^{0,0}
+\sqrt{\frac{2}{45}}I^{1,-1}+\sqrt{\frac{2}{45}}I^{-1,1}\big],
\end{eqnarray}
where  $I^{0,0}$, $I^{1,-1}$ and $I^{-1,1}$ are
same as  in Eqs. (\ref{I00}) and (\ref{I11}) \cite{Zhang2007}.


\subsubsection{$\mathbf{1^{-}(2^{3}S_{1})}$ $\mathbf{c\bar{s}}$\textbf{ state}}

The harmonic oscillator wave function of the
$1^{-}(2^{3}S_{1})$ $c\bar{s}$ state is given by
\begin{eqnarray}
\phi^{n=2;L=0}({{\mathbf{k}_1,\mathbf{k}_2}}) &=&
\frac{1}{\sqrt{4\pi}}\bigg (\frac{4R^3}{\sqrt{\pi}}\bigg)^{{1}/{2}}
\sqrt{\frac{2}{3}} \bigg[\frac{3}{2} - \frac{R^2}{4}
 ({{\mathbf{k}_1-\mathbf{k}_2})}^2 \bigg]\exp\Big{[}-{1\over 8}({{\mathbf{k}_1-\mathbf{k}_2})}^2
 R^2\Big{]}.
 \nonumber
\end{eqnarray}
The general decay amplitude of $c\bar{s}(2^{3}S_{1})\to
1^{-}+0^{-}$ is given  as

\begin{eqnarray}
\mathcal{M}(c\bar{s}(2^{3}S_{1})\to 1^{-}+0^{-}
)&=&\frac{\gamma\sqrt{8E_{D^*_s}E_{D^*}E_{K}}}{\sqrt{3}} \bigg(
\frac{1}{ \sqrt{18}} I^{0,0}_{0,0}  \bigg),
\end{eqnarray}

where
\begin{eqnarray} I^{0,0}_{0,0}&=& - \sqrt{\frac{1}{2}}
\frac{|\mathbf{k}_{D^*}|
    \pi^{1/4}R_{D^*_s}^{3/2}R_{D^*}^{3/2}R_{K}^{3/2}}
{(R_{D^*_s}^{2}+R_{D^*}^{2}+R_{K}^{2})^{5/2}}             \Bigg\{
-6(R_{D^*_s}^{2}+R_{D^*}^{2}+R_{K}^{2})(1+\Upsilon)+R_{D^*_s}^2  \bigg[
4+20\Upsilon\nonumber\\&&+\mathbf{k}_{D^*}^2
(R_{D^*_s}^{2}+R_{D^*}^{2}+R_{K}^{2}) (-1+\Upsilon)^2(1+\Upsilon) \big]
 \Bigg\}
\exp\bigg[-\frac{\mathbf{k}_{D^*}^{2}R_{D^*_s}^{2}(R_{D^*}^{2}+R_{K}^{2})}{8(R_{D^*_s}^{2}+R_{D^*}^{2}+R_{K}^{2})}\bigg].
\label{express-3}
\end{eqnarray}

In above equations, the variables $R_{D^*_s}$, $R_{D^*}$ and $R_{K}$  represent the radius of respective mesons.
  The parameter $\alpha$ is  taken to be 1 \cite{Zhang2007} and the  universal parameter  $\gamma$ is taken as 6.9 \cite{Zhang2007,Godfrey1996}. In addition, the parameter $\Upsilon$ is expressed as $\Upsilon=\frac{R_{D^*_s}^2}{R_{D^*_s}^2+R_{D^*}^2+R_{K}^2}$. 

\section{Numerical Results and Discussions}
\label{sec:3}

\begin{table}
\begin{tabular}{|c|c|c|c|c|}

\hline
$g_{\sigma N}$  & $g_{\zeta N }$  &  $g_{\delta N }$  &
$g_{\omega N}$ & $g_{\rho N}$ \\

\hline 
10.56 & -0.46 & 2.48 & 13.35 & 5.48  \\

\hline 
$\sigma_0$ (MeV)& $\zeta_0$(MeV) & $\chi_0$(MeV)  & $d$ & $\rho_0$ ($\text{fm}^{-3}$)  \\ 
\hline 
-93.29 & -106.8 & 409.8 & 0.064 & 0.15  \\ 
\hline

\hline 
$m_\pi $(MeV) &$ m_K$ (MeV)&$ f_\pi$(MeV)  & $f_K$(MeV) & $g_4$ \\ 
\hline 
139 & 494 & 93.29 & 122.14 & 79.91  \\ 
\hline

\hline
$k_0$ & $k_1$ & $k_2$ & $k_3$ & $k_4$  \\ 
\hline 
2.53 & 1.35 & -4.77 & -2.77 & -0.218  \\ 
\hline

\end{tabular}
\caption{Values of different parameters used in the chiral model calculations.} \label{ccc}
\end{table} 

  In this section, we will discuss our observations of the magnetic field induced mass spectra of  vector $({D^{*}}^+,{D^{*}}^0)$ and axial-vector $(D^+_1,D^0_1)$  mesons. As discussed earlier, the light quark condensates and gluon condensates have been calculated by using chiral SU(3) model and the different parameters used in model are tabulated in Table \ref{ccc}. In present work, the mass and decay constant of $D$ mesons has been calculated using QCDSR. In this calculation, the value of charm quark mass $m_c$, running coupling constant $\alpha_s$, coupling constant $g_{DNH}$ and constant $\lambda$ are approximated to be  1.3 GeV, 0.45, 3.86 and 0.5, respectively \cite{Kumar2015,Wang2015}. In addition, the vacuum masses of 
${D^{*}}^+$, ${D^{*}}^0$,  ${D^+_1}$ and  ${D^0_1}$ mesons  
 are taken as 2.010, 2.006, 2.423 and 2.421 GeV,  respectively.    
 The vacuum values of the decay constant for vector and axial-vector mesons are taken as 0.270 and 0.305 GeV, respectively. In QCDSR, we use proper Borel window to ensure the observed properties have minimum variation in their value for a range of Borel parameter.  The Borel window for masses of $({D^{*}}^+,{D^{*}}^0)$ and  $(D^+_1,D^0_1)$ are taken as  (4.5-5.5)  and (6-7) GeV$^2$, respectively, whereas the range of Borel window for decay constant of $({D^{*}}^+,{D^{*}}^0)$ and   $(D^+_1,D^0_1)$ are taken as (2-3)    and (7-9) GeV$^2$, respectively.  We have divided our discussion into two parts.

  \subsection{In-medium mass and shift in decay constant of vector $D^*$ and axial-vector $D_1$ mesons}
 \label{subsec:3.1}
 
\begin{figure}
\includegraphics[width=16cm,height=21cm]{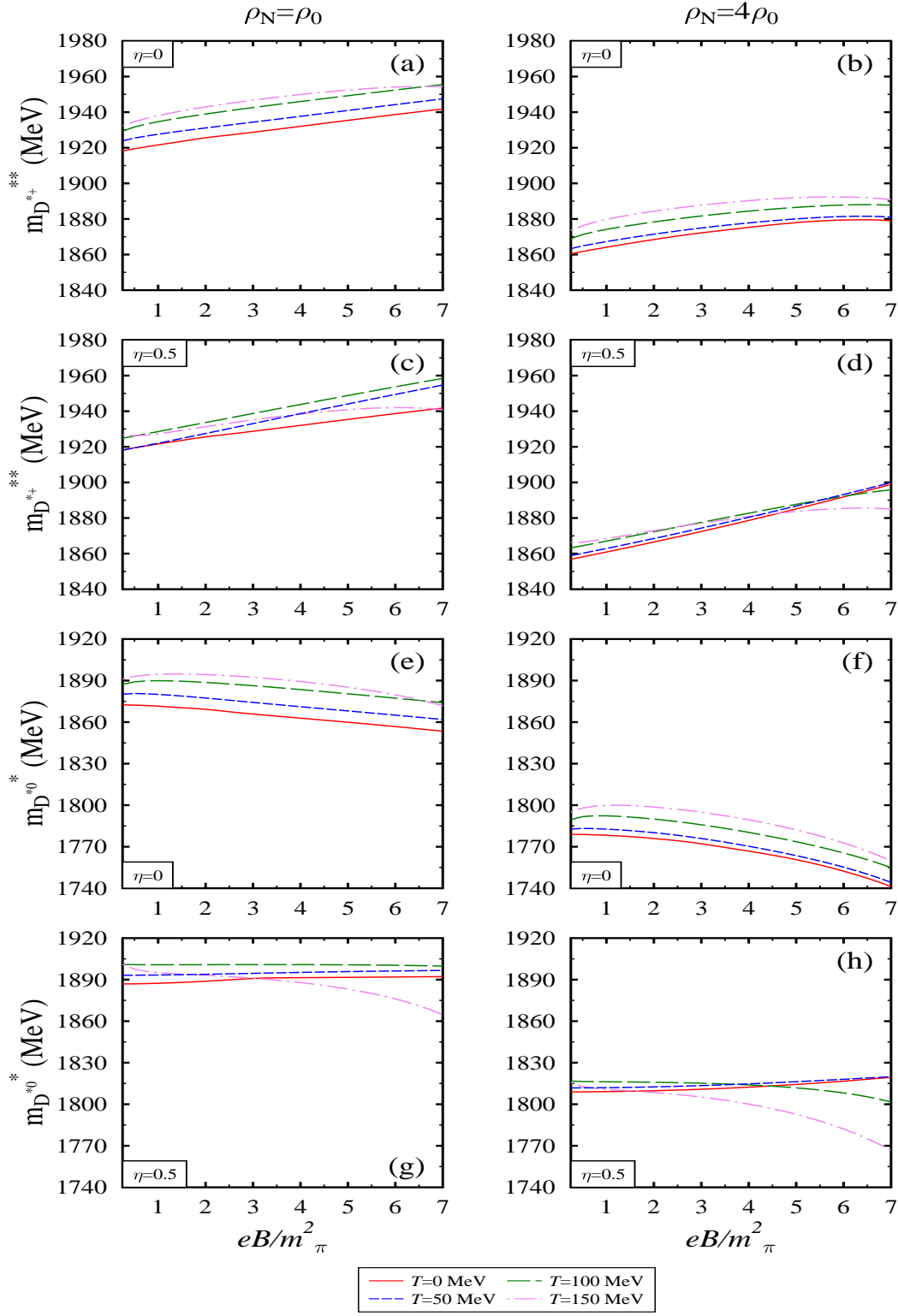}
\caption{(Color online) In the above figure the variation of in-medium mass of vector $D^*(D^{*^+},D^{*^0})$ mesons. }
\label{msv}
\end{figure}

\begin{figure}
\includegraphics[width=16cm,height=21cm]{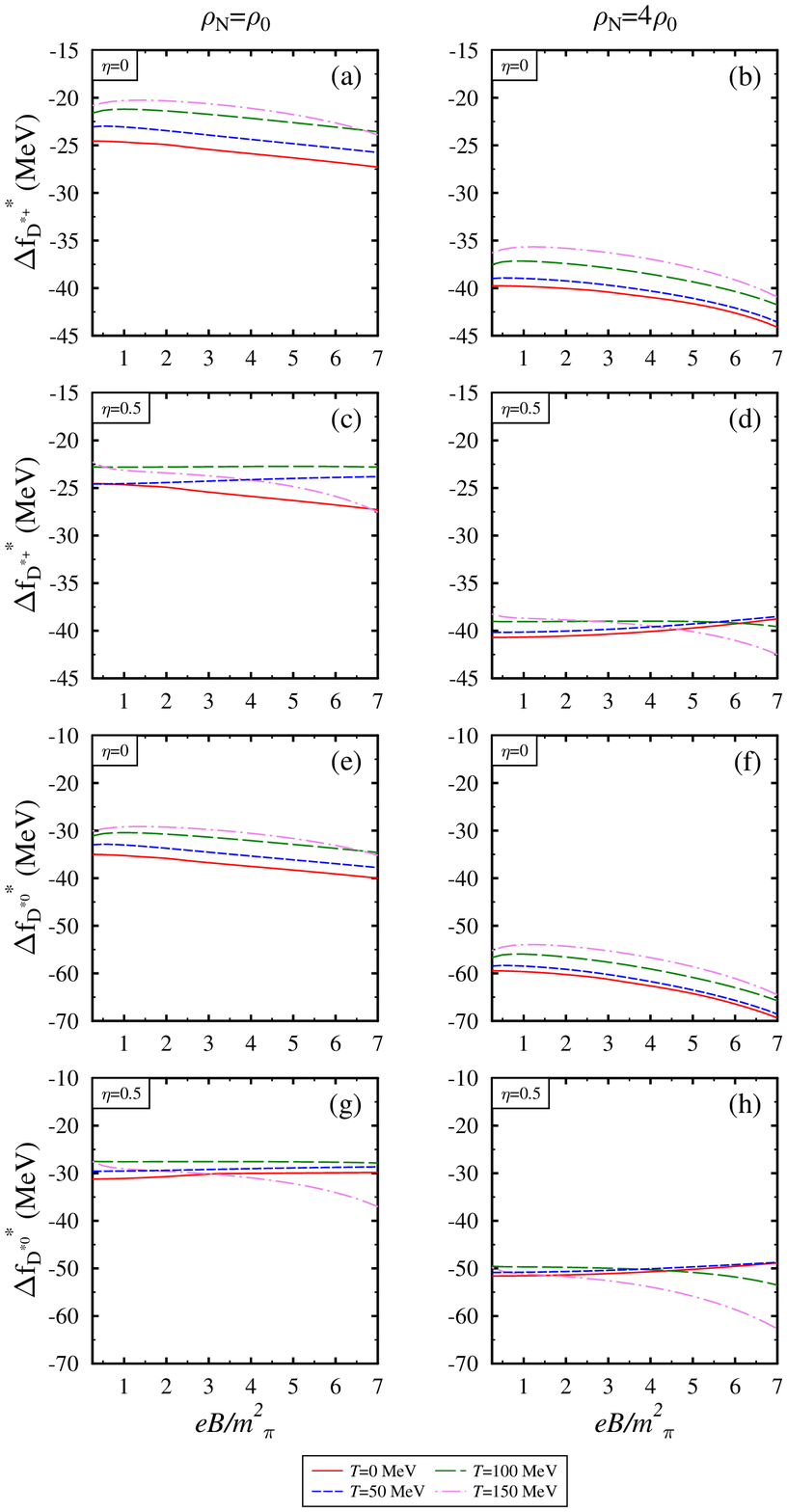}
\caption{(Color online) In the above figure the variation of shift in decay constant of vector $D^*(D^{*^+},D^{*^0})$ mesons.}
\label{fdv}
\end{figure}

In \cref{msv} (\cref{fdv}), we represent the in-medium mass (decay constant) of $D^{*+}$ and $D^{*0}$ mesons in hot and dense isospin asymmetric nuclear medium as a function of magnetic field. The numerical values of in-medium masses and decay constants are given in \cref{tablems,tablefd}, respectively. We observe that, for any constant value of temperature $T$, isospin asymmetric parameter $\eta$ and nuclear density $\rho_N$ of the medium, the finite magnetic field $e B / m^2 _\pi $ of the medium  causes an enhancement (drop) in the mass (shift in decay constant) of $D^{*+}$ meson, whereas drop (drop) in the mass (shift in decay constant) of $D^{*0}$ meson.  For example, in symmetric nuclear medium, at temperature $T=0$ and nuclear density $\rho_N$=$\rho_0$, the masses (shift in decay constant) of $D^{*+}$ and $D^{*0}$ mesons are observed to be 1928 (-25.43) and 1865 (-36.73) MeV, respectively, at finite magnetic field $i.e.$, $eB$ = $3m^2 _\pi$. On increasing the magnetic field to $eB$ = $5 m^2 _\pi$, the above values shift to 1935 (-26.31) and 1859 (-38.29) MeV, respectively. Further, we notice that the mass of $D^{*+}$ is more sensitive to the presence of the magnetic field of the medium, as compared to $D^{*0}$ mesons, and this is because of the Landau effect. In this effect, the charged particle couple with the magnetic field and starts revolving in circular levels called Landau levels. This quantization modifies the scalar and vector density of the nucleons and in turn the quark and gluon condensates \cite{Kumar2020}.   Further, for any constant value of magnetic field, isopin asymmetric parameter and temperature of the medium  the masses and decay constants of these mesons decrease  as a function of nuclear density of the medium. For example, in cold symmetric nuclear medium and at four times nuclear saturation density, i.e., $\rho_N$ = 4$\rho_0$ the masses (shift in decay constant) of   $D^{*+}$ and   $D^{*0}$ mesons are observed to be 1872 (-40.41) and 1772 (-61.23)  MeV, respectively at $eB$ = 3$m^2 _\pi$. Also at $eB$ = 5$m^2 _\pi$ the above values shift to  1877 (-41.63) and 1760 (-64.25) MeV, respectively. Moreover, in finite magnetic field the effect of the finite temperature on the in-medium mass and decay constants of  $D^{*+}$ and $D^{*0}$ mesons is opposite to that of nuclear density of the medium. For example, in symmetric nuclear medium, at finite magnetic field of the medium, i.e., $eB$ = 3$m^2 _\pi$ (5$m^2 _\pi$) the masses of   $D^{*+}$ and $D^{*0}$ mesons are observed as 1942 (1949) and 1886 (1880) MeV, at temperature $T$ = 100 MeV and $\rho_N=\rho_0$, respectively. Likewise at  $\rho_N=4\rho_0$, the above values modify to 1881 (1886) and  1785 (1773) MeV, respectively. We can compare these values at zero temperature situation, as discussed earlier. 

  Apart from this, finite isospin asymmetry of the medium causes splitting between the in-medium masses (decay constants) of vector $D^*$  mesons. For example, in an isospin asymmetric medium, $\eta$ = 0.5, $\rho_N=\rho_0$  and $eB$ = 3$m^2 _\pi$ the values of the masses (decay constants) of $D^{*+}$ and $D^{*0}$ mesons observed to be  1923 (-25.10) and 1890 (-30.17) MeV at $\rho_N$ = $\rho_0$, temperature $T$ = 0. Likewise, at $eB$ = 5$m^2 _\pi$ the above value change to 1940 (-24.9) and 1891 (-29.96) MeV, respectively. Clearly these values are different from the masses and decay constants of   $D^{*+}$ and $D^{*0}$ mesons as discussed for the symmetric matter situation. The presence of magnetic field helps to increase the asymmetry effect of the medium. The scalar/vector density of the proton has direct magnetic field dependence whereas for neutron there is no such dependence. The increase in magnetic field increase the inequality between scalar/vector density of the neutron and proton and hence give rise to the crossover asymmetry effects \cite{Kumar2020}.

\begin{figure}
\includegraphics[width=16cm,height=21cm]{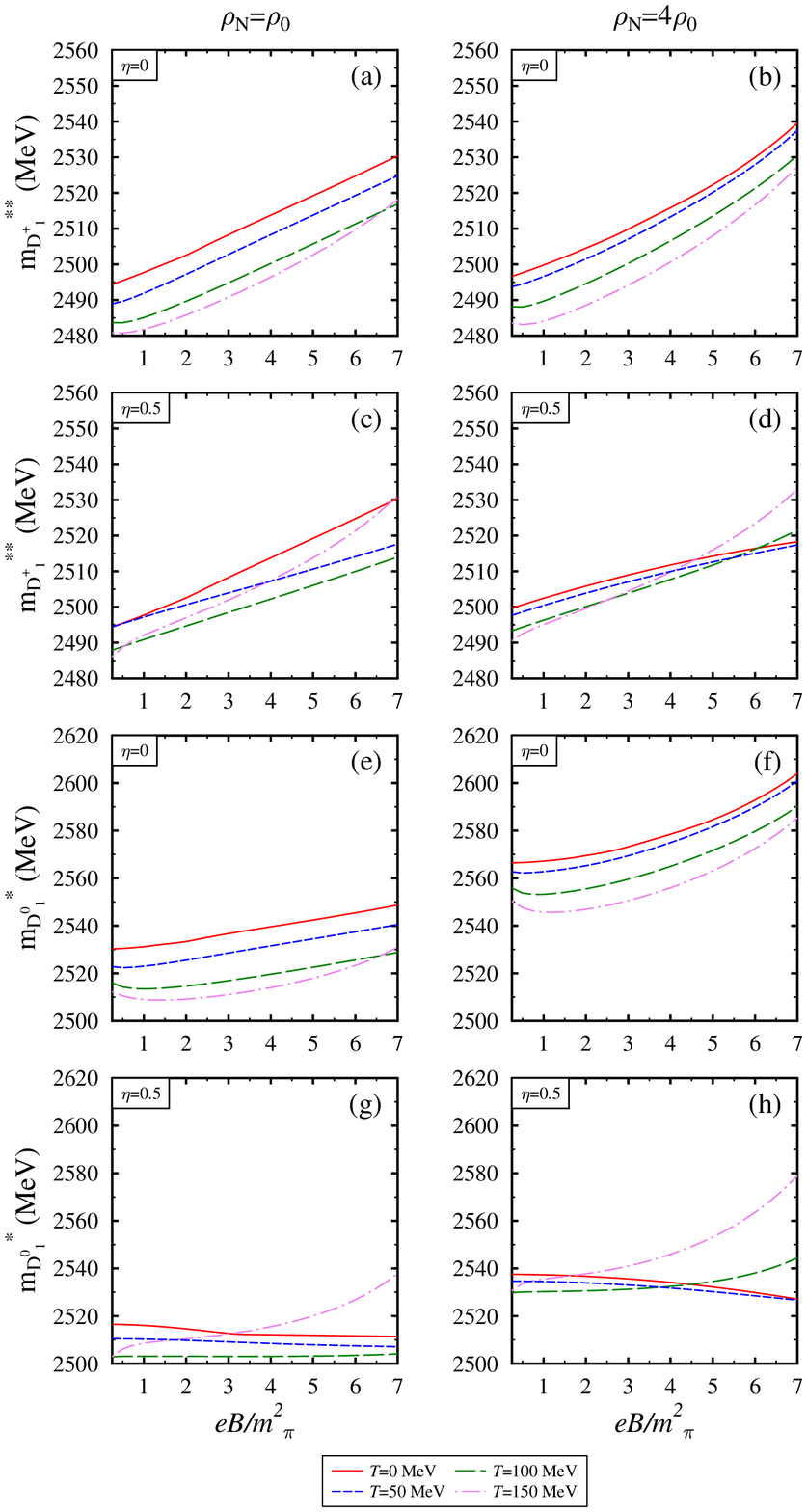}
\caption{(Color online) In the above figure the variation of in-medium mass of axial-vector $D_1(D^+_1, D^0_1)$ mesons. }
\label{msa}
\end{figure}

 In \cref{msa}, we observe an enhanced in-medium mass of axial-vector $D_1$ meson in magnetised nuclear matter as compared to non-magnetised nuclear matter. In particular, in cold symmetric nuclear medium and at nuclear saturation density, the values of the mass of $D_1^+$ ($D_1^0$) meson observed as 2508 (2536) and 2519 (2542) MeV, respectively, at $eB$ = 3$m^2 _\pi$ and $eB$ = 5$m^2 _\pi$. In addition to magnetic field, the finite baryonic density of the medium also causes an increase in the values of the masses of $D_1^+$ ($D_1^0$) meson. For instance, in cold symmetric nuclear medium and $\rho_N$ = 4$\rho_0$, the above values  modifies  to 2509 (2573) and 2522 (2584) MeV, respectively. Similar to vector mesons, the impact of the finite temperature of the medium on the in-medium mass of axial-vector mesons is also opposite to that of the density and magnetic field. For example, at $T$=100 MeV,  $eB$ = 3$m^2 _\pi$ and $eB$ = 5$m^2 _\pi$, the masses of $D_1^+$ ($D_1^0$) meson are observed to be 2494 (2516) and 2505 (2522) MeV, respectively, at  $\eta$=0, $\rho_N$=$\rho_0$. 
On the other hand, finite isospin asymmetric parameter causes further modification in the masses of $D_1^+$ and $D_1^0$ mesons. As discussed earlier, the crossover effects represents the behaviour of quark and gluon condensates which in turns reflects the medium modification of the scalar fields in the nuclear medium \cite{Kumar2020,Kumar2019}. 

\begin{figure}
\includegraphics[width=16cm,height=21cm]{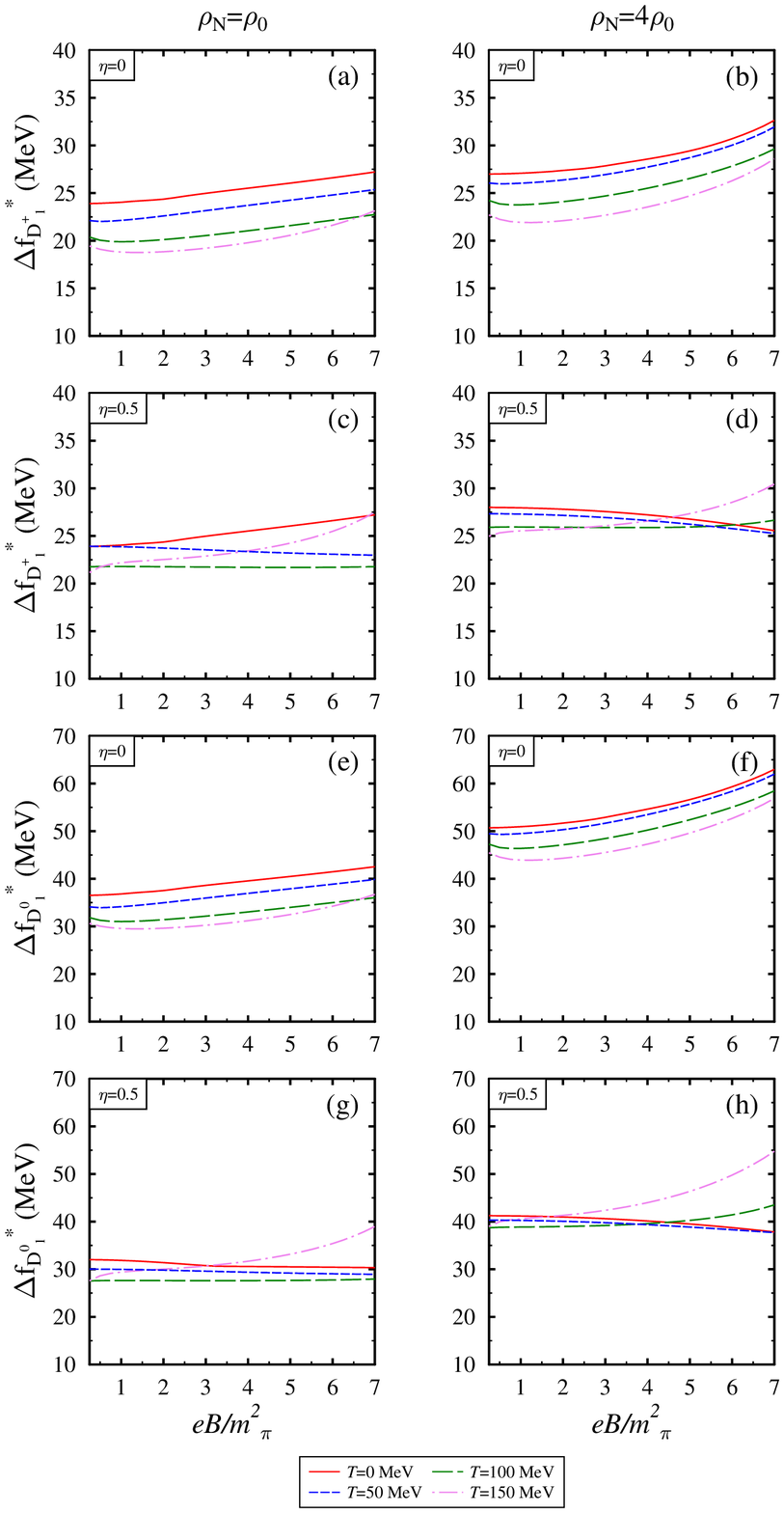}
\caption{(Color online) In the above figure the variation of  shift in decay constant of axial-vector $D_1(D^+_1, D^0_1)$ mesons. }
\label{fda}
\end{figure}

 In \cref{fda}, we have shown the shift in decay constants of axial-vector $D$ mesons. As similar to the masses, the shift in the  in-medium decay constants of $D_1^+$ and $D_1^0$ mesons increase with the increase in magnetic field and these are shown particularly, at $eB$ =3$m^2 _\pi$ and $eB$ = 5$m^2 _\pi$, the values of the shift in decay constants of $D_1^+$ ($D_1^0$) mesons are observed as 24.97 (38.61) and 26.04 (40.48) MeV, respectively, at $T$=0, $\eta$=0 and $\rho_N$=$\rho_0$. Likewise, at $\rho_N$=4$\rho_0$  the above values shift to 27.85 (52.50) and 29.42 (56.62) MeV, respectively. Whereas, the role of finite temperature is to decrease the shift in the decay constants. For example, in hot nuclear medium, say at $T$ = 100 MeV,  $\rho_N$=$\rho_0$ and $\eta$=0, the values of decay constants of $D_1^+$ ($D_1^0$) mesons are observed to be 20.53 (32.15) and 21.58 (33.99) MeV, at $eB$ = 3$m^2 _\pi$ and $eB$ = 5$m^2 _\pi$, respectively. The finite 
isospin asymmetry of the medium also causes the splitting in the in-medium decay constants of $D_1^0$ and $D_1^+$ mesons.  In the present investigation, we found opposite  behaviour of the masses and decay constants of vector $D^*$ and axial-vector $D_1$ mesons. By ignoring the additional mass shift due to Landau effect, we observe that the value of vector $D^*$ meson's mass and shift in decay constant decreases  as a function of magnetic field whereas for axial-vector $D_1$ mesons these in-medium attributes increase. This happens because of the negative sign with the term $\frac{m_{c}\langle\bar{q}q\rangle_N}{2}$ (see Eq. (\ref{coef})), of the Borel transformed equation. This causes negative and positive values of the scattering length for  $D^* N$ and $D_1 N$, scattering, respectively in the presence of finite magnetic field \cite{Wang2015,Kumar2015}. In the absence of magnetic field, the in-medium mass of heavy vector and axial-vector have also been studied in the  literature \cite{Kumar2014,Kumar2015,Wang2015}. However, in our knowledge, no work is available till date which study the impact of magnetic field on the vector and axial-vector $D$ mesons.

\begin{figure}
\includegraphics[width=16cm,height=21cm]{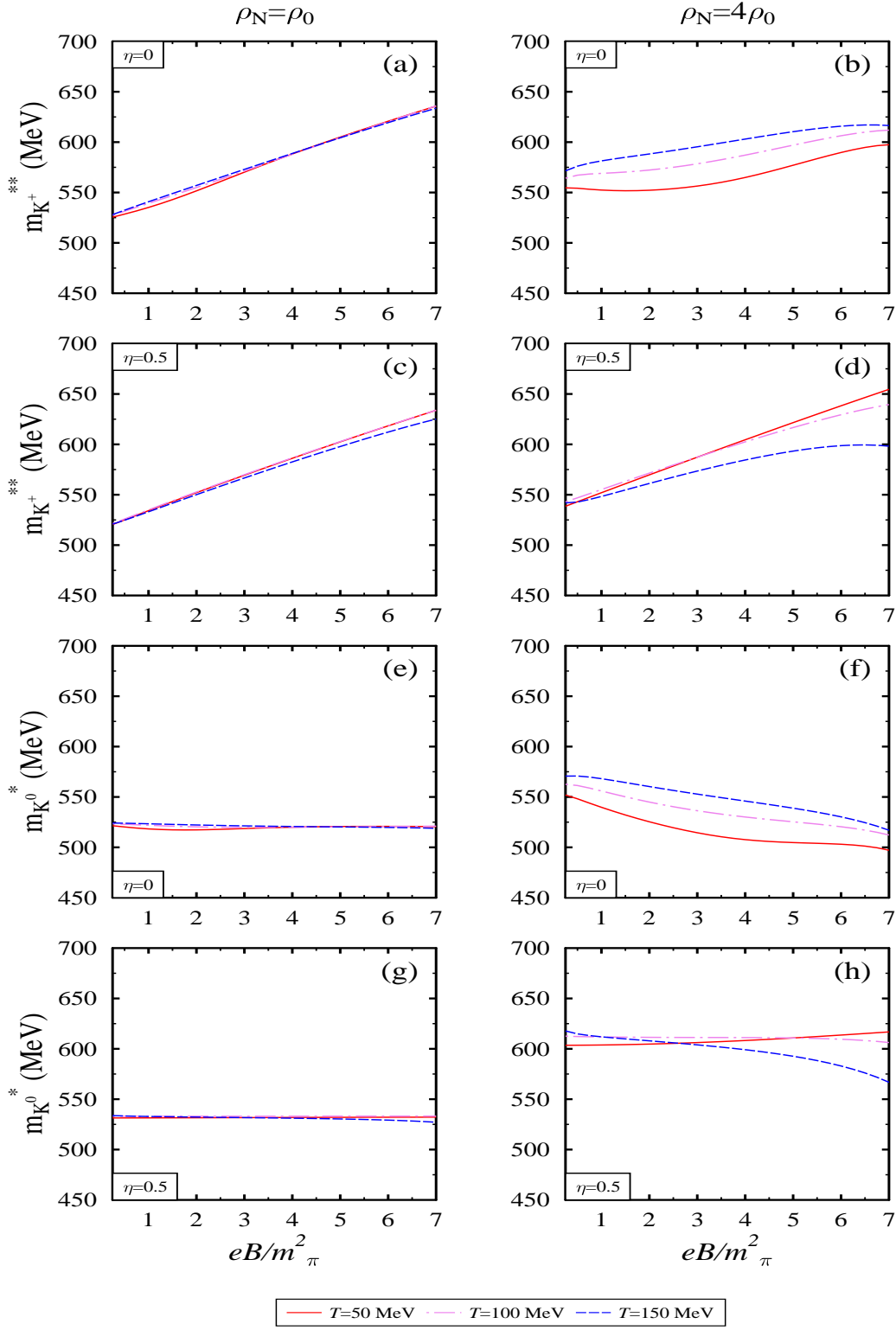}
\caption{(Color online) In the above figure we plot the variation of in medium mass of $K^+$ and $K^0$ mesons.}
\label{msk}
\end{figure}

In  \cref{msk}, the in-medium masses of $K^+$ and $K^0$ mesons are also plotted in hot, dense and  asymmetric magnetized nuclear matter. As a function of magnetic field, we observe positive shift in the mass of $K^+$ mesons in both low and high density of nuclear matter. On the other hand, for neutral $K^0$ meson, we observe negative shift in mass  with respect to magnetic field. The negative shift is appreciable in the high density regime.   This is because, the self energy of  kaons (see Eq. (\ref{selfk})) directly depends upon the scalar and vector density of the nucleons. The charged $K^+$ meson largely depend upon scalar and vector density of proton whereas the uncharged $K^0$ depends upon neutron's scalar and vector density. In addition to this, in the presence of magnetic field the  $K^+$ meson experience additional positive shift in the mass whereas uncharged $K$ meson does not experience any additional shift. It may be noted that, in the presence of magnetic field the value of isoscalar $\delta$  field no longer remains zero  in symmetric nuclear matter \cite{Kumar2020} which also induce additional asymmetry effect. In our findings at nuclear saturation density, $\eta$=0, $T=50$ MeV and $eB$ = 5$m^2 _\pi$, the medium mass of $K^+$ ($K^0$) is observed as 605 (520) MeV . At $\eta=0.5$, and same other parameters the value changes to 602 (535) MeV.  The Weinberg Tomozawa term (first term of the self energy), leads to increase in the mass of kaon and the second term give attractive contributions. The remaining range terms gives repulsive (third term) and attractive contributions ($d_1$ and $d_2$ terms). Furthermore, we observe a little effect of the temperature on the mass of kaons in low density regime, however this effect is appreciable in the high density regime.  In Ref.\cite{Mishra2019}, author studied the  in-medium mass of $K$ and $\bar K$ meson in the  asymmetric magnetized nuclear matter using chiral SU(3) model at zero temperature. In this article, they calculated the in-medium mass of $K$ and $\bar K$ meson with and without taking the account of anomalous magnetic moment of nucleons. Now, we will utilize the  in-medium mass of $D^*$   and $K$ meson  to calculate the magnetic field induced decay width of $D^*_s$ meson.

 \begin{table}
 \begin{tabular}{|c|c|c|c|c|c|c|c|c|c|}
\hline
& & \multicolumn{4}{c|}{$\eta$=0}    & \multicolumn{4}{c|}{$\eta$=0.5}   \\
\cline{3-10}
&$eB/{{m}_{\pi}^2}$ & \multicolumn{2}{c|}{T=0} & \multicolumn{2}{c|}{T=100 }& \multicolumn{2}{c|}{T=0}& \multicolumn{2}{c|}{T=100 }\\
\cline{3-10}
&  &$\rho_0$&$4\rho_0$ &$\rho_0$  &$4\rho_0$ & $\rho_0$ &$4\rho_0$&$\rho_0$&$4\rho_0$ \\ \hline 
$ m^{**}_{{D^*}^+}$& 3&1928& 1872& 1942 &1881&1923&1872&1938&1877 \\ \cline{2-10}
&5&1935&1877  &1949  & 1886 & 1940 &1885 &1948 & 1887 \\ \cline{1-10}
$m^*_{{D^*}^0}$&3&1865 & 1772 & 1886 &1785 &1890 &1810 & 1888 &1815 \\  \cline{2-10}
&5&1851 &1760 &1880 &1773 & 1891 &1814 & 1884&1811 \\  
 \cline{1-10}
$ m^{**}_{D_1^+}$&3&2508 & 2509 & 2494 &2500 &2506&2508 & 2498 &2503\\  \cline{2-10}
&5&2519 &2522 &2505 &2513 & 2514 &2514 &2506&2511\\  \hline
$ m^*_{D_1^0}$&3&2536 & 2573 & 2516 &2559 &2512 &2535 & 2503 &2531 \\  \cline{2-10}
&5&2542 &2584 &2522 &2571 & 2511 &2532 &2503&2534\\  \cline{1-10}
\end{tabular}
\caption{In above table, we tabulated the values of in-medium masses   of ${D^*}^+$, ${D^*}^0$, $D_1^+$ and $D_1^0$ mesons (in units of MeV).}
\label{tablems}
\end{table}

 \begin{table}
\begin{tabular}{|c|c|c|c|c|c|c|c|c|c|}
\hline
& & \multicolumn{4}{c|}{$\eta$=0}    & \multicolumn{4}{c|}{$\eta$=0.5}   \\
\cline{3-10}
&$eB/{{m}_{\pi}^2}$ & \multicolumn{2}{c|}{T=0} & \multicolumn{2}{c|}{T=100 }& \multicolumn{2}{c|}{T=0}& \multicolumn{2}{c|}{T=100 }\\
\cline{3-10}
&  &$\rho_0$&$4\rho_0$ &$\rho_0$  &$4\rho_0$ & $\rho_0$ &$4\rho_0$&$\rho_0$&$4\rho_0$ \\ \hline 
$ \Delta f^{*}_{{D^*}^+}$& 3& -25.43&-40.41& -21.73 &-37.89&-25.10&-40.35&-22.76&-38.79\\ \cline{2-10}
&5&-26.31  &-41.63  & -22.61 & -39.34 &-24.90 &-39.21& -22.72 & -39.03 \\ \cline{1-10}
$ \Delta f^*_{{D^*}^0}$&3&-36.73 & -61.23 & -31.30 &-57.64 &-30.17 &-51.09 & -27.55 &-49.95 \\  \cline{2-10}
&5&-38.29 &-64.25 &-32.90 &-60.85 & -29.96 &-50.18 & -27.59&-50.83 \\  
 \cline{1-10}
$  \Delta f^{*}_{D_1^+}$&3&24.97 & 27.85 & 20.53 &24.69 &24.52 &27.55&21.73 & 25.88  \\  \cline{2-10}
&5&26.04 &29.42 &21.58 &26.53 & 24.31 &26.76 &21.68&25.93 \\  \hline
$  \Delta f^*_{D_1^0}$& 3&38.61&52.90 &32.15&48.45&30.76&40.62&27.60 &39.19\\ \cline{2-10}
&5&40.48  &56.62  &33.99 & 52.39&30.51 &39.50 & 27.65 & 40.26 \\ \cline{1-10}
\end{tabular}
\caption{In above table, we tabulated the values of in-medium shift in decay constant   of ${D^*}^+$, ${D^*}^0$, $D_1^+$ and $D_1^0$ mesons (in units of MeV).}
\label{tablefd}
\end{table}

 \subsection{ Decay width of $D^*_s$ $\rightarrow$ $D^*$ $K$  meson}
 \label{subsec:3.2}

Now, we will calculate the in-medium decay width of the $D^*_s$ meson into $D^*$ and $K$ mesons. The $D^*_s$ meson is expected to have one of the quantum state  $1^{-}(1^{3}{D}_{1})$, $3^{-}(1^{3}{D}_{3})$ and $1^{-}(2^{3}{S}_{1})$ with mass of 2860 or 2715 MeV \cite{Zhang2007,Chhabra2017}. In the present work, we have calculated the decay width for three different cases for each value of mass. The results will be compared with the theoretical findings and empirical values, which will help in the future to assign a quantum state to parent meson. In the present investigation,  we have neglected the medium modifications of the parent meson.  To the best of our knowledge, no study has been done to study the in-medium properties of $D^*_s$(2715) and $D^*_s$(2860) meson in the presence or absence of magnetic field in nuclear medium. In \cref{tabledw1,tabledw2}, we have tabulated the   decay width of  $D^*_s$(2715) and $D^*_s$(2860) mesons, respectively at  $R_A$=2.94 GeV$^{-1}$ for $1^{-}(1^{3}D_{1})$, $3^{-}(1^{3}D_{3})$ states   and 3.2 GeV$^{-1}$ for $1^{-}(2^{3}S_{1})$ state.
 The values of $R_{D^*}$ and $R_{K}$  are taken as 2.70 and 2.17 GeV$^{-1}$, respectively \cite{Chhabra2017,Zhang2007}

\begin{figure}
\includegraphics[width=16cm,height=16cm]{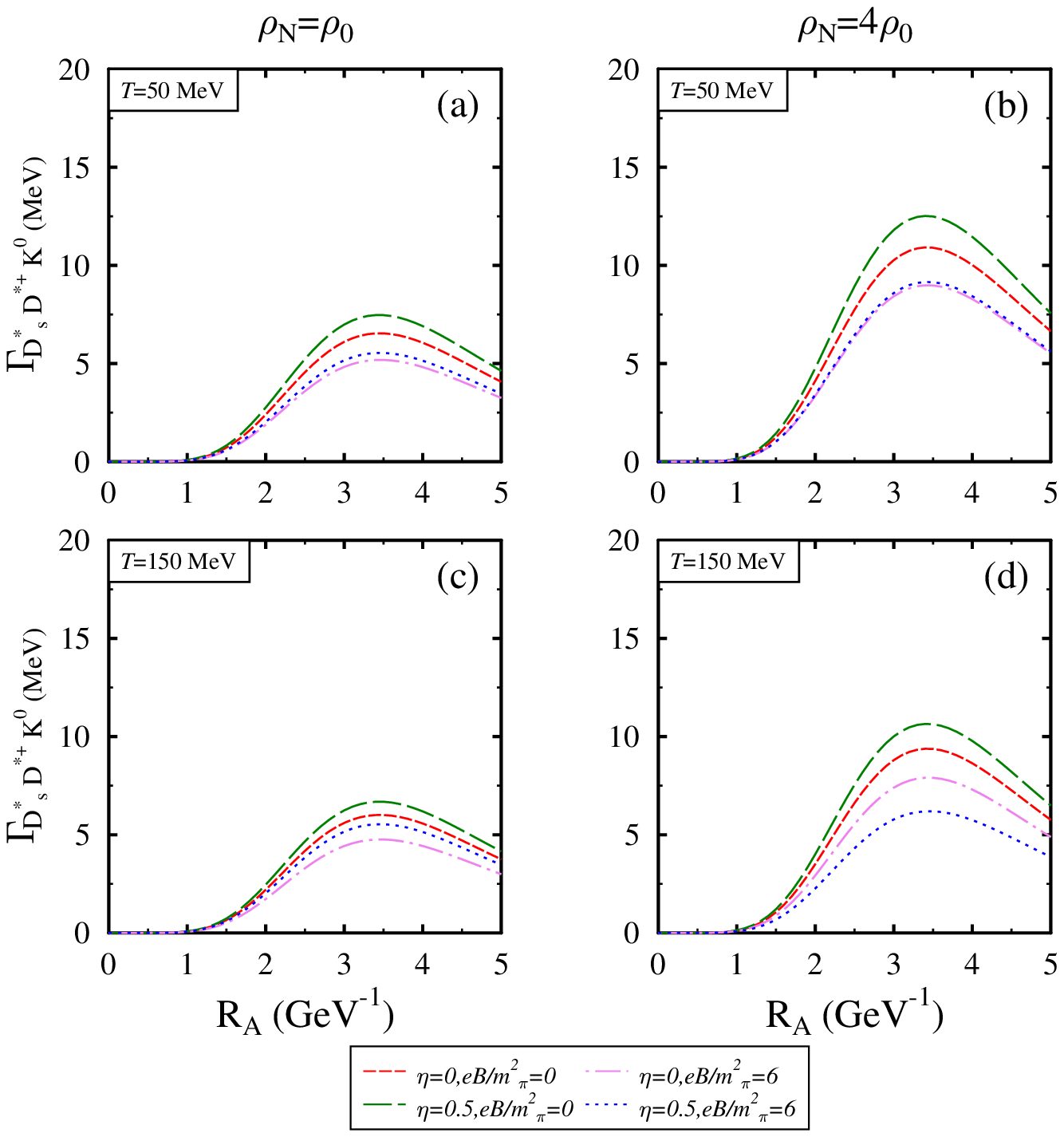}
\caption{(Color online) The in-medium decay width of $D^*_s(2715)$ in $1^{-}(1^{3}D_{1})$ spectroscopic state is plotted with respect to $R_A$ for different conditions of the medium. }
\label{27151d3}
\end{figure}
\begin{figure}
\includegraphics[width=16cm,height=16cm]{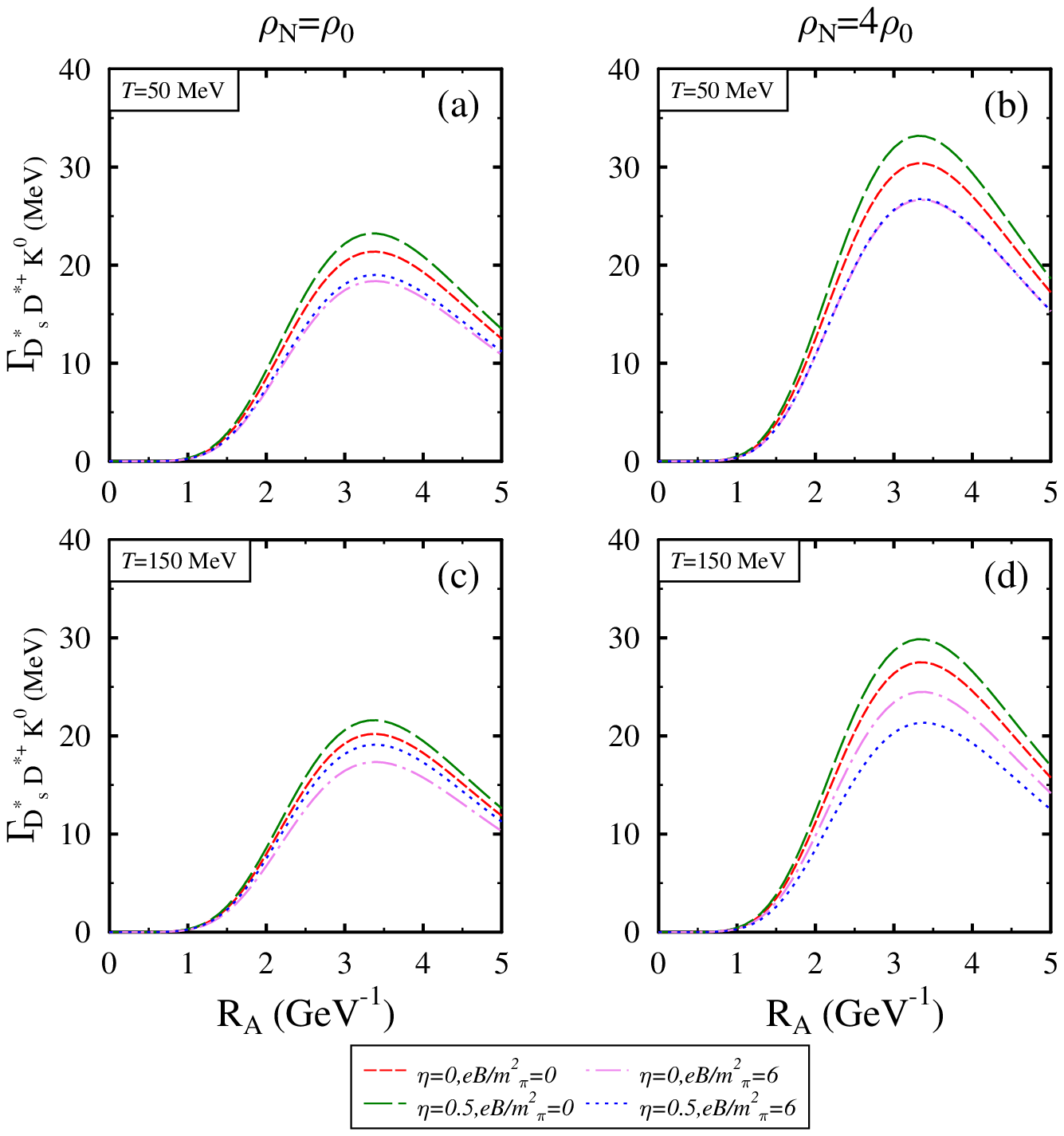}
\caption{(Color online) The in-medium decay width of $D^*_s(2860)$ in $1^{-}(1^{3}D_{1})$ spectroscopic state is plotted with respect to  $R_A$ for different conditions of the medium. }
\label{28601d3}
\end{figure}

In \cref{27151d3}, for a given value of temperature, density, magnetic field and isospin asymmetry, we have plotted the decay width of  $D^*_s$(2715) meson in  $1^{-}(1^{3}{D}_{1})$ state as a function of  $R_A$. We observed zero value of decay width  upto $R_A$ $\sim$1 GeV$^{-1}$. The decay width increase upto $R_A$ $\sim$3.35 GeV$^{-1}$ and then decreases with more increase in $R_A$.  The decay width shows similar trend for different values of medium parameters. Inclusion of magnetic field effects show significant impact on the decay width.  Further, the value of decay width increases when we move from low density to high density. We also observed that the temperature effects are more appreciable in the asymmetric nuclear matter as in the current framework the scalar and vector densities consist of Fermi distribution functions.
  Now, in \cref{28601d3}, in the same quantum state, when we change the mass of $D^*_s$ meson from 2715 MeV to 2860 MeV, we observe similar behaviour of decay width with respect to different medium parameters. The only difference in $D^*_s$(2860) is that we observe a large value of decay width for each parameter as compared to $D^*_s$(2715) case and the maxima of decay width shifted near $R_A$ $\sim$3.2 GeV$^{-1}$.

\begin{figure}
\includegraphics[width=16cm,height=16cm]{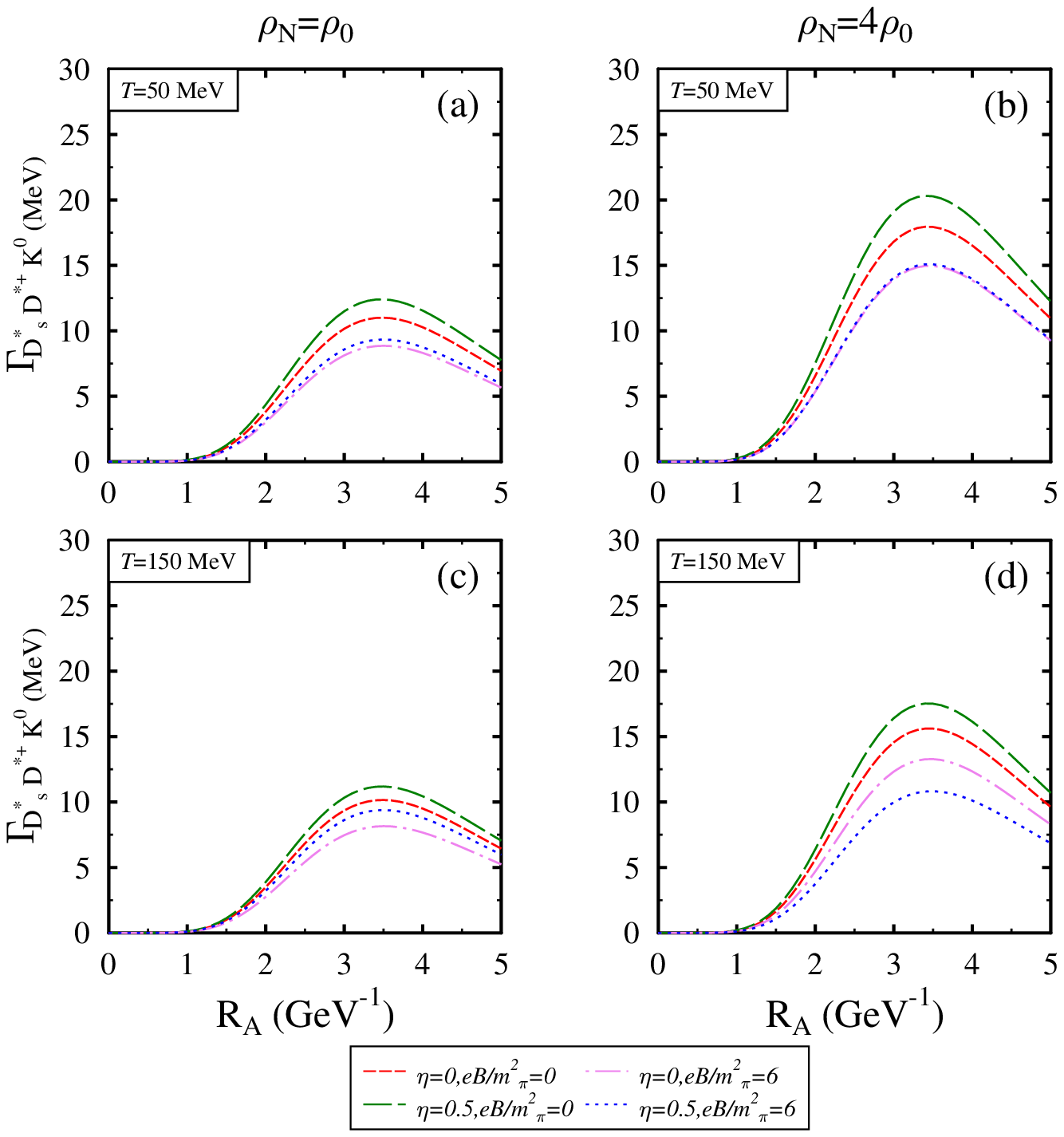}
\caption{(Color online) The in-medium decay width of $D^*_s(2715)$ in $3^{-}(1^{3}D_{3})$ spectroscopic state is plotted with respect to  $R_A$ for different conditions of the medium. }
\label{271531d3}
\end{figure}

\begin{figure}
\includegraphics[width=16cm,height=16cm]{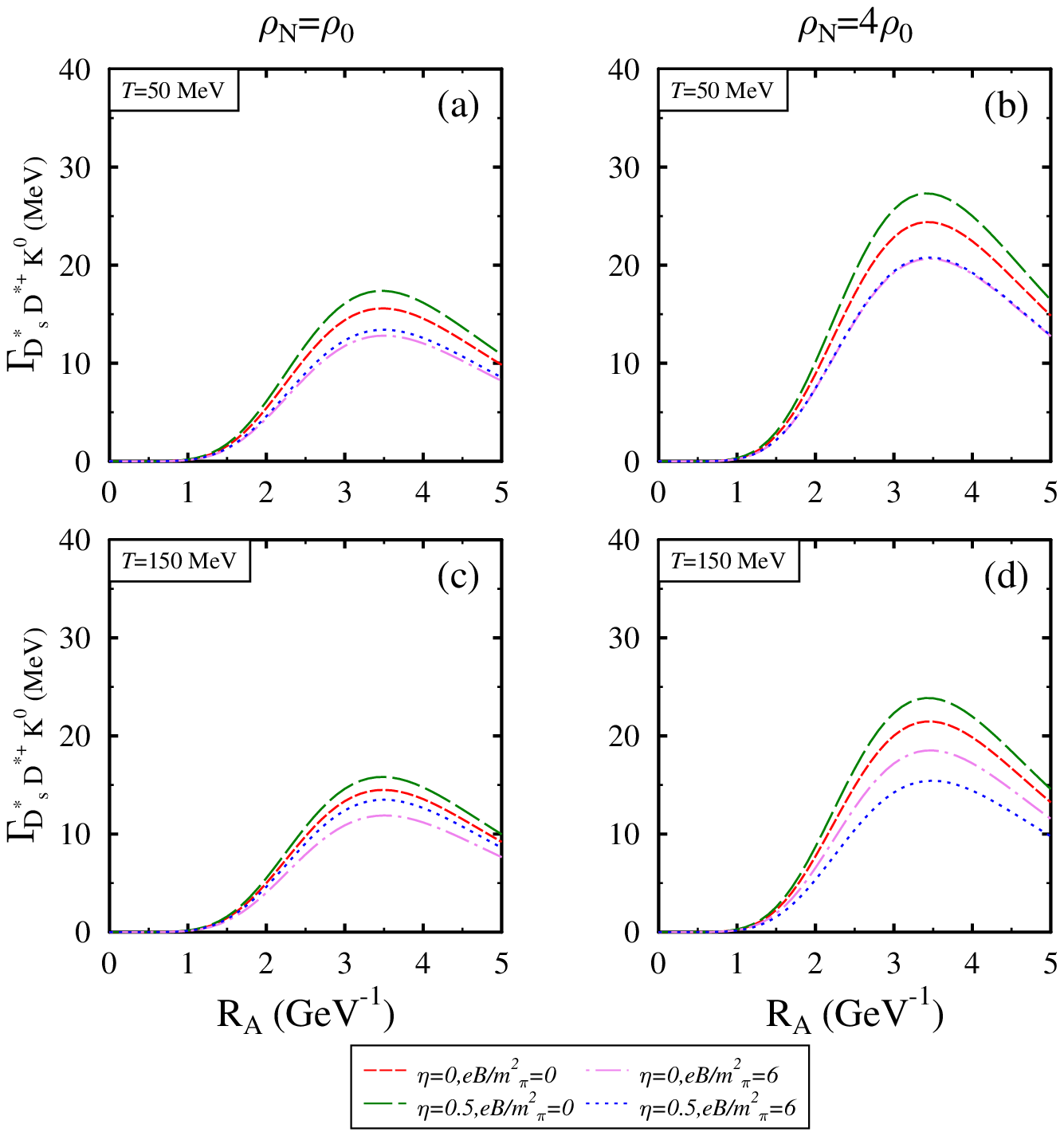}

\caption{(Color online) The in-medium decay width of $D^*_s(2860)$ in $3^{-}(1^{3}D_{3})$ spectroscopic state is plotted with respect to  $R_A$ for different conditions of the medium. }
\label{286031d3}
\end{figure}

\begin{figure}
\includegraphics[width=16cm,height=16cm]{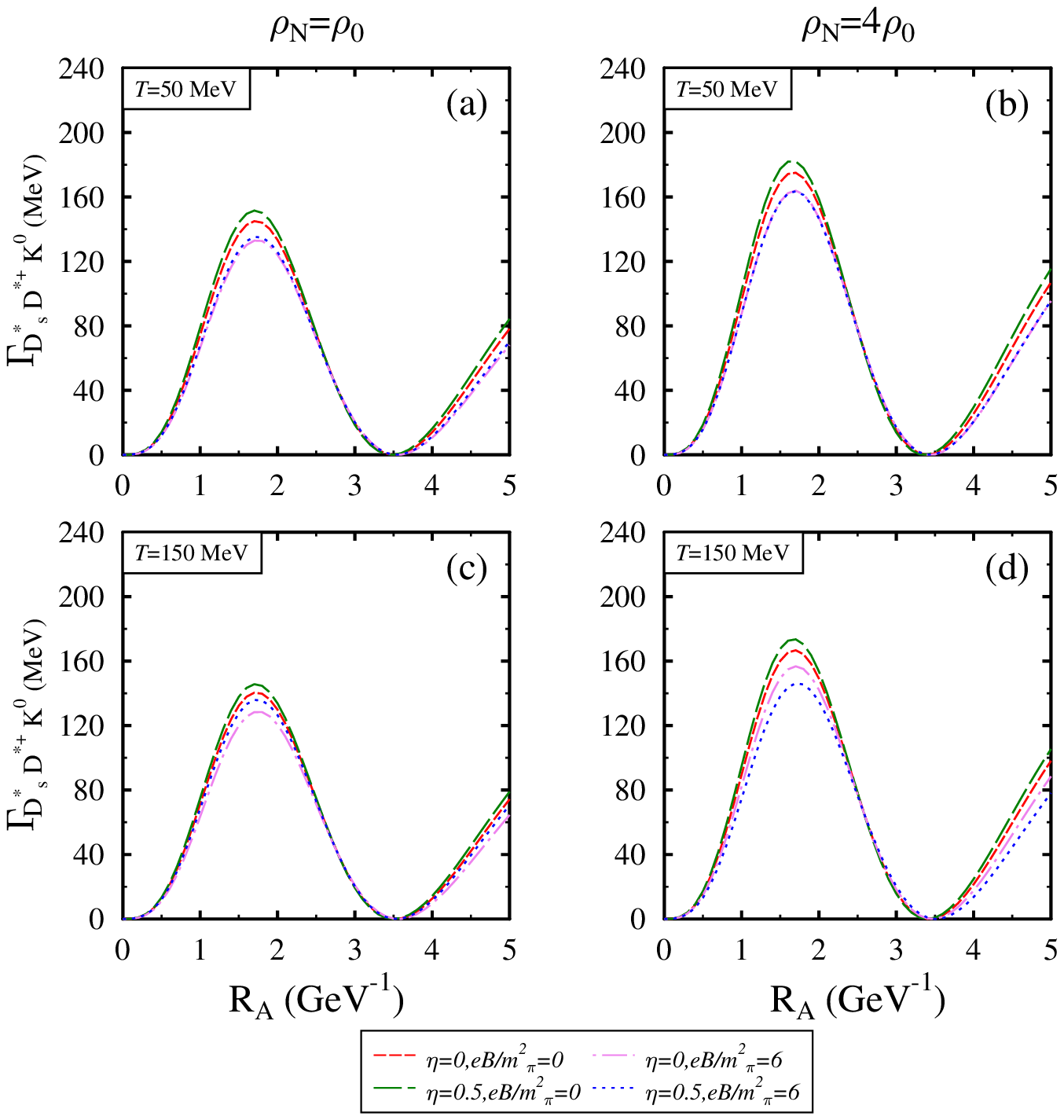}
\caption{(Color online) The in-medium decay width of $D^*_s(2715)$ in $1^{-}(2^{3}S_{1})$ spectroscopic state is plotted with respect to  $R_A$ for different conditions of the medium.}
\label{27152s1}
\end{figure}

\begin{figure}
\includegraphics[width=16cm,height=16cm]{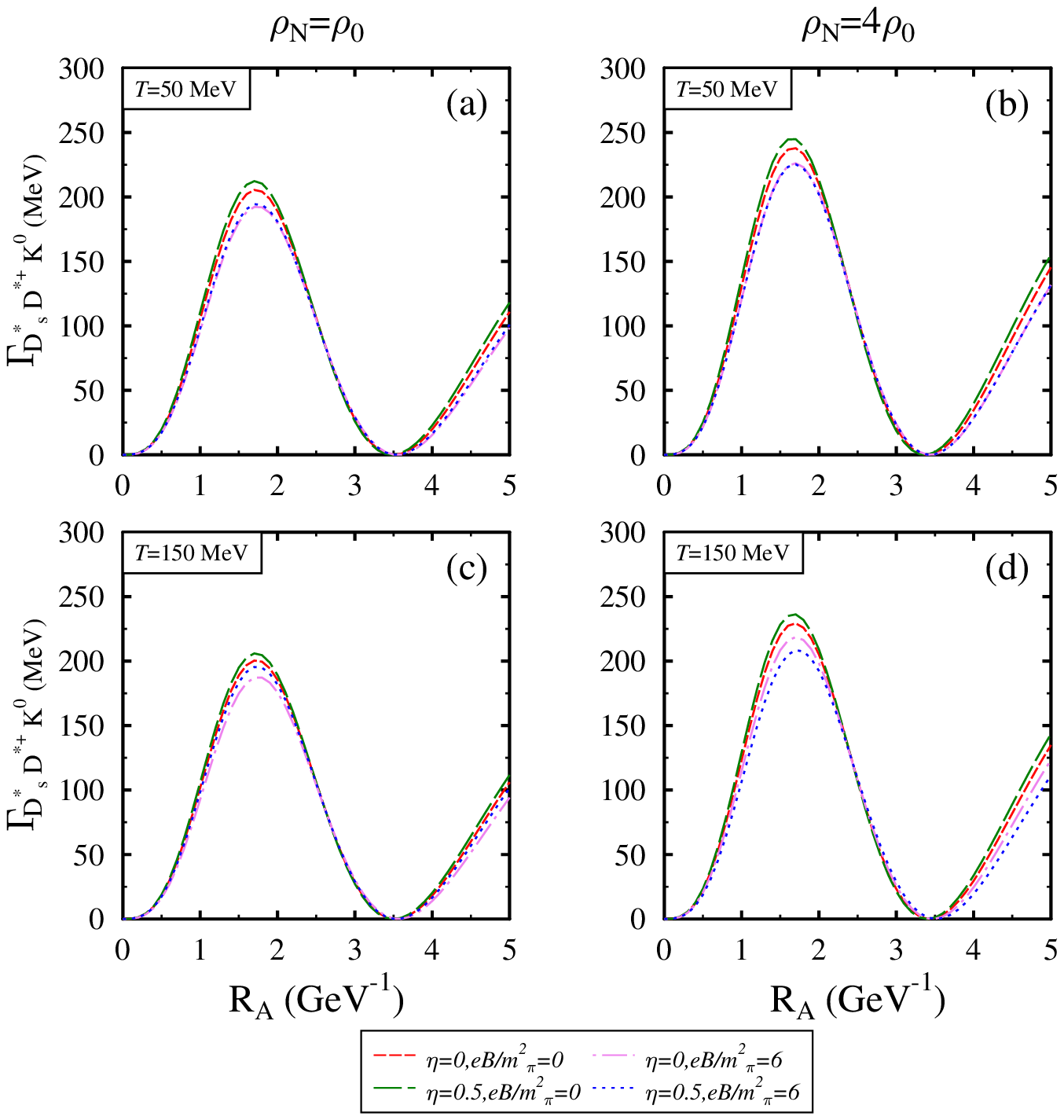}

\caption{(Color online) The in-medium decay width of $D^*_s(2860)$ in $1^{-}(2^{3}S_{1})$ spectroscopic state is plotted with respect to $R_A$ for different conditions of the medium.}
\label{28602s1}
\end{figure}

In \cref{271531d3,286031d3}, we have plotted the decay width of $D^*_s$ meson in $3^{-}(1^{3}{D}_{3})$ state as a function of $R_A$ in the presence of magnetic field, asymmetry, temperature and density for mass values 2715 and 2860 MeV, respectively. We observe that at first decay width increase upto certain value of $R_A$ and it starts decreasing with further increase in $R_A$. We observe appreciable effect of density on the decay width of the $D^*_s$ meson. The effect of magnetic field is more in the high density regime than the low density regime. Also, the asymmetry effects are more visible in the high temperature regime. This is due to  the fact that the decay width inversely relates to the in-medium mass. If mass  of $K$ and $D^*$ meson  decrease more in the medium then we observe more increase in the decay width.   Furthermore, in the same spectroscopic state, we observed large value of decay width for $D^*_s$(2860) as compared to $D^*_s$(2715) meson. Furthermore, the decay width of $D^*_s$(2715) and $D^*_s$(2860) meson in $1^{-}(2^{3}{S}_{1})$ quantum state is calculated and shown in \cref{27152s1,28602s1}, respectively. In this spectroscopic state, we observe large value of decay width of $D^*_s$ meson in both mass state. In the former case, we observe  the decay width first increase with the increase in $R_A$ and after reaching at certain value $R_A$ $\sim$ 1.65 GeV$^{-1}$, it starts decreasing following a node at $R_A$ $\sim$ 3.35 GeV$^{-1}$ and then it again increases for further values of $R_A$.  In this case, we observe appreciable effect of magnetic field in the presence of density and temperature. The formula of decay width consist of two terms, polynomial and exponential. This behaviour reflects the interplay between these two parts. 
Furthermore, as discussed earlier, we also observe appreciable asymmetric effects in the high temperature regime. Moreover, for the $D^*_s$(2860) meson, we observe high value of decay width, however the other medium effects remains same. The vacuum  decay width of the decay mode $D^*_s$ (in different spectroscopic states) into $\rightarrow$ $D^*$ $K$ and various other modes has been calculated in literature \cite{Zhang2007}. But the present theoretical and empirical data is not enough to assign a quantum state to the $D^*_s$ meson. A more detailed theoretical and experimental study of the in-medium properties of $D^*_s$ meson is needed to understand its correct quantum state.

 \begin{table}
\begin{tabular}{|c|c|c|c|c|c|c|c|c|c|c|}
\hline
    & & \multicolumn{4}{c|}{$\eta$=0} & \multicolumn{4}{c|}{$\eta$=0.5}  \\
\cline{3-10}
 & & \multicolumn{2}{c|}{T=50} & \multicolumn{2}{c|}{T=150 } &\multicolumn{2}{c|}{T=50} & \multicolumn{2}{c|}{T=150 }\\
\cline{3-10}
 $D^*_s$(2715) $\rightarrow$ $D^*$ $K$ &$eB/{{m}_{\pi}^2}$&  $\rho_0$ &$4\rho_0$&$\rho_0$&$4\rho_0$ &   $\rho_0$ & $4\rho_0 $ & $\rho_0$ & $4\rho_0$\\ \hline
$1^{-}(1^{3}D_{1})$&0
&5.9&5.3&9.5&6.86&5.84&5.28&5.26&6.31\\  \cline{2-10} 
 & 6 &4.84&4.4&7.47&7.47&4.22&4.67&5.63&4.03 \\  \cline{1-10}

$3^{-}(1^{3}D_{3})$   &0 &9.8&8.91&15.48&11.67&9.77&8.87&9.29&10.94 \\   \cline{2-10} 
 & 6 &8.10&7.39&12.38&12.38&7.15&7.86&9.71&7.24\\  \cline{1-10} 
 
$1^{-}(2^{3}S_{1})$  &0&7.4&7.99&4.38&6.64&7.52&8.07&8.27&7.22\\  \cline{2-10} 
  & 6 &8.46&8.88&6.05&6.05&9.09&8.65&7.84&9.49 \\  \cline{1-10} 
 
\hline

\end{tabular}

\caption{In the above table, we listed the value of magnetic field induced decay width (MeV) of $D^*_s(2715)$ mesons into vector $D^*$ and $K$ mesons for different parameters of the medium at particular value of $R_A$.}
\label{tabledw1}
\end{table}

\begin{table}
\begin{tabular}{|c|c|c|c|c|c|c|c|c|c|c|}
\hline
   &   & \multicolumn{4}{c|}{$\eta$=0} & \multicolumn{4}{c|}{$\eta$=0.5}  \\
\cline{3-10}
&   & \multicolumn{2}{c|}{T=50} & \multicolumn{2}{c|}{T=150 } &\multicolumn{2}{c|}{T=50} & \multicolumn{2}{c|}{T=150 }\\
\cline{3-10}
 $D^*_s$(2860)  $\rightarrow$ $D^*$ $K$ &$eB/{{m}_{\pi}^2}$&  $\rho_0$ &$4\rho_0$&$\rho_0$&$4\rho_0$ &   $\rho_0$ & $4\rho_0 $ & $\rho_0$ & $4\rho_0$\\ \hline
$1^{-}(1^{3}D_{1})$ &0&5.9&5.33&9.52&6.88&5.84&5.2&5.26&6.31\\  \cline{2-10} 
  & 6 &4.8&4.41&7.47&7.47&4.22&4.67&5.63&4.03 \\  \cline{1-10}

$3^{-}(1^{3}D_{3})$   &0 &9.8&8.91&11.67&15.48&9.77&8.87&9.29&10.94\\   \cline{2-10} 
  & 6 &8.1&7.39&12.38&12.38&7.15&7.86&9.71&7.24\\  \cline{1-10} 
 
$1^{-}(2^{3}S_{1})$  &0&7.4&7.99&4.38&6.64&7.52&8.07&8.27&7.22\\  \cline{2-10} 
   & 6 &8.46&8.88&6.05&6.05&9.09&8.65&7.84&9.49 \\  \cline{1-10} 
 
\hline

\end{tabular}

\caption{In the above table, we listed the maxima of magnetic field induced decay width (MeV) of $D^*_s(2860)$ mesons into vector $D^*$ and $K$ mesons for different parameters of the medium at particular value of $R_A$. }
\label{tabledw2}
\end{table}

%
%
%
%
%
%
%
%

\section{Conclusion}
\label{sec:4}

To conclude, we studied the medium modification of the masses and shift in decay constant of vector and axial-vector $D$ meson under the influence of strong magnetic field. We also studied the effect of isospin asymmetry, temperature and density alongside the magnetic field.   The in-medium mass of $K$ meson is calculated by using the chiral model and to calculate the in-medium mass of $D$ mesons, we used the unified approach of QCD Sum Rule (QCDSR) and chiral SU(3) model. The scalar (vector) density of nucleons modifies differently due to the presence of magnetic field and isospin asymmetry. The magnetic field interacts differently with charged proton and give rise to Landau effect.  The non linear coupled equations of different mesons fields are solved by incorporating the magnetic field induced density which in turn generates the light quark and gluon condensates. We found appreciable effects of  strong magnetic field on the charged vector ${D^*}^+$ and axial-vector $D^+_1$ mesons whereas for uncharged $D$ mesons, the effects were less appreciable. We calculated negative (positive) mass shift for vector (axial-vector) uncharged $D$ mesons and for charged vector and axial-vector $D$ meson, we found positive shift in mass with respect to the magnetic field. The density and temperature effects on these mesons were also appreciable. The isospin asymmetry effects are suppressed by the Landau quantization for the charged meson case whereas for neutral mesons it shows crossover behaviour. By utilising the in-medium mass, we calculated the magnetic field induced decay width of $D^*_s$ meson decaying to $D^{*+}$ and $K^0$ meson via $^3P_0$ model and observed prominent modifications in the decay width of $D^*_s$ mesons in $1^{-}(1^{3}{D}_{1})$, $3^{-}(1^{3}{D}_{3})$ and $1^{-}(2^{3}{S}_{1})$ spectroscopic states.  The observed decay probability will be compared with the experimental results of future experiments such as CBM, PANDA, J-PARC and NICA.



\centering
\section*{Acknowledgement}

One of the author, (R.K)  sincerely acknowledge the support towards this work from Ministry of Science and Human Resources Development (MHRD), Government of India via Institute fellowship under National Institute of Technology Jalandhar.

\end{document}